\documentclass[11pt]{article}

\usepackage[table,dvipsnames]{xcolor}
\usepackage{multirow}

\usepackage[]{ACL2023}

\usepackage{times}
\usepackage{latexsym}

\usepackage[T1]{fontenc}

\usepackage[utf8]{inputenc}

\usepackage{microtype}

\usepackage{inconsolata}

\usepackage{amsmath}
\usepackage{graphics}
\usepackage{layouts}
\usepackage{graphicx}
\usepackage{subfigure}
\usepackage{cleveref}
\usepackage{booktabs}
\usepackage{caption}
\usepackage{array}
\usepackage{tabularray}
 \usepackage{enumitem}
\usepackage{arydshln}
\usepackage{xspace}
\usepackage{adjustbox}
\usepackage{xcolor}

\usepackage{listings}

\definecolor{lightgray}{gray}{0.9}
\newcommand{\sysname}[0]{\textit{EASE}\xspace}

\usepackage{amsmath}

\colorlet{punct}{red!60!black}
\definecolor{background}{HTML}{EEEEEE}
\definecolor{delim}{RGB}{20,105,176}
\colorlet{numb}{magenta!60!black}

\lstdefinelanguage{json}{
    basicstyle=\footnotesize\ttfamily,
    numberstyle=\scriptsize,
    stepnumber=1,
    numbersep=4pt,
    showstringspaces=false,
    breaklines=true,
    frame=lines,
    backgroundcolor=\color{background},
    literate=
     *{0}{{{\color{numb}0}}}{1}
      {1}{{{\color{numb}1}}}{1}
      {2}{{{\color{numb}2}}}{1}
      {3}{{{\color{numb}3}}}{1}
      {4}{{{\color{numb}4}}}{1}
      {5}{{{\color{numb}5}}}{1}
      {6}{{{\color{numb}6}}}{1}
      {7}{{{\color{numb}7}}}{1}
      {8}{{{\color{numb}8}}}{1}
      {9}{{{\color{numb}9}}}{1}
      {:}{{{\color{punct}{:}}}}{1}
      {,}{{{\color{punct}{,}}}}{1}
      {\{}{{{\color{delim}{\{}}}}{1}
      {\}}{{{\color{delim}{\}}}}}{1}
      {[}{{{\color{delim}{[}}}}{1}
      {]}{{{\color{delim}{]}}}}{1},
}

\title{\sysname: An \underline{\textit{E}}asily-Customized \underline{\textit{A}}nnotation \underline{\textit{S}}ystem Powered by \\\underline{\textit{E}}fficiency Enhancement Mechanisms}

\author{Naihao Deng$^{*\diamondsuit}$ \quad 
Yikai Liu$^{*\diamondsuit}$ \quad
Mingye Chen$^{*\diamondsuit}$ \quad
Winston Wu$^{\diamondsuit}$ \quad \\
\bf Siyang Liu$^{\diamondsuit}$ \quad
\bf Yulong Chen$^{\dagger}$ \quad
\bf Yue Zhang$^{\dagger}$ \quad
\bf Rada Mihalcea$^{\diamondsuit}$ \quad \\
$^\diamondsuit$ University of Michigan \quad $^\dagger$Westlake University \\
{\tt \{dnaihao,yikai.liu,mingyech\}@umich.edu}}

\UseRawInputEncoding

\begin{document}
\maketitle
\begin{abstract}

\def\thefootnote{*}\footnotetext{Naihao Deng, Yikai Liu, and Mingye Chen contributed equally to the manuscript.}

The performance of current supervised AI systems is tightly connected to the availability of annotated datasets. 
Annotations are usually collected through annotation tools, which are often designed for specific tasks and are difficult to customize.
Moreover, existing annotation tools with an active learning mechanism often only support limited use cases.
To address these limitations, we present {\bf \sysname}, an \textbf{\textit{E}}asily-Customized \textbf{\textit{A}}nnotation \textbf{\textit{S}}ystem Powered by \textbf{\textit{E}}fficiency Enhancement Mechanisms.
\sysname provides modular annotation units for building customized annotation interfaces and also provides multiple back-end options that suggest annotations using (1) multi-task active learning; (2) demographic feature based active learning; (3) a prompt system that can query the API of large language models.
We conduct multiple experiments and user studies to evaluate our system's flexibility and effectiveness. Our results show that our system can meet the diverse needs of NLP researchers and significantly accelerate the annotation process.

\end{abstract}

\section{Introduction}


The surge of data-driven methods has accentuated the demand for faster data annotation, with various tasks and applications requiring customized data collection solutions. 
Recent years have witnessed a significant increase in the number of proposed benchmarks \citep{zhang2021ai}, and a myriad of tasks and datasets have been emerging to address the resource bottleneck of AI applications \citep{deng2009imagenet,bowman-etal-2015-large,rajpurkar-etal-2016-squad,swayamdipta-etal-2020-dataset,mazumder2022dataperf}, 
suggesting the necessity for data annotation systems that allow fast and economic data construction under diverse needs.
Many existing annotation systems either are designed for specific tasks, or do not adequately assist the annotator in efficiently providing annotations.
In this paper, we propose {\bf \sysname}, an \textbf{\textit{E}}asily-Customized \textbf{\textit{A}}nnotation \textbf{\textit{S}}ystem Powered by \textbf{\textit{E}}fficiency Enhancement Mechanisms, aiming to assist researchers with high-volume data annotation in more effective ways. \sysname has two main advantages over existing annotation interfaces:



(1) \sysname is flexible, supporting a wide range of customization options.
Existing annotation tools mostly support annotations for specific tasks \citep{li2021fitannotator} or have fixed templates that are difficult to customize to meet the specific needs of researchers. In contrast, we define a series of minimum front-end modular units, which researchers can specify using JSON. \sysname will then render the corresponding interface. 
This allows researchers to build the annotation interface according to their intents while exempting them from the tedious work of coding the interface from scratch.

\begin{figure}[t]
    \centering
    \includegraphics[width=0.95\linewidth]{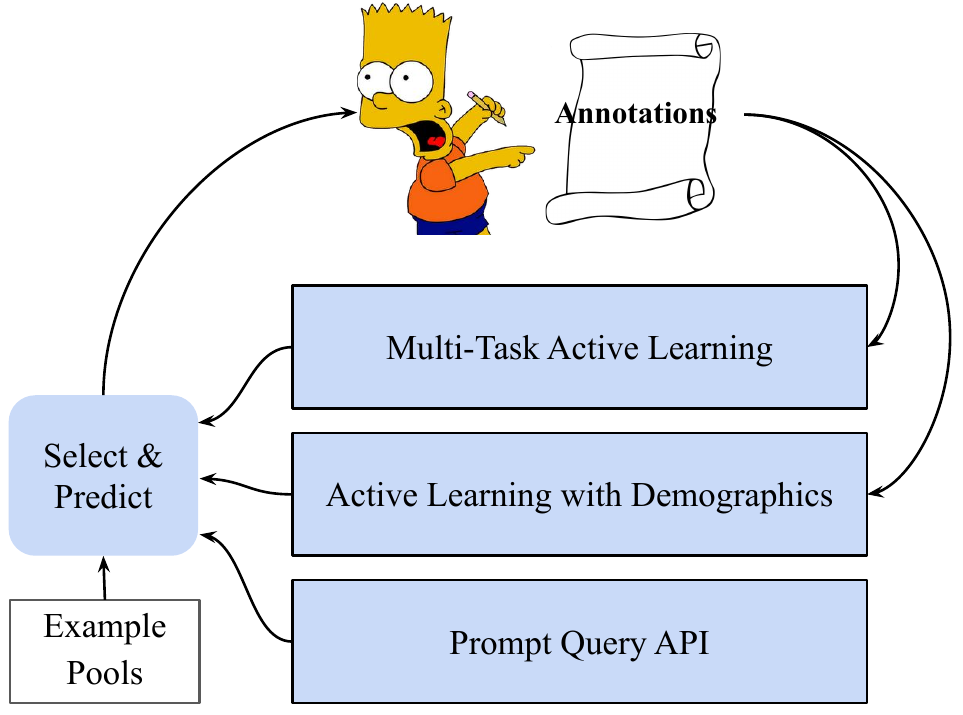}
    \caption{System workflow for \sysname. We provide a customizable front end interface (\Cref{sec: front-end}) with 3 back-end options of multi-task active learning (\Cref{sec: mtal-backend}), active learning with demographic features (\Cref{sec: al-demographics-backend}), prompt query API (\Cref{sec: prompt-backend}).}
    \label{fig:concept-diagram}
\end{figure}

\begin{table*}[ht]
    \small
	\centering
    \begin{adjustbox}{width=\linewidth}
		\begin{tabular}{@{}ccccc@{}}
			\toprule
			Task Family & Task & Dataset & Input Form & Annotation Form \\
			\midrule
            \multirow{6} * {\it Text Classification}  & Ontology Classification & DBpedia-14~\cite{zhang2015character} & Question & Multi-choice \\
               & Fact Checking & FEVER~\cite{thorne-etal-2018-fever} & Text & Multi-choice \\
               &  Sentiment Analysis & Restaurants~\cite{pontiki2016semeval} & Question & Scores \\
               &  Textual Entailment & SNLI~\cite{bowman-etal-2015-large} & Statements & Multi-choice \\
                & \textit{Syntactic Analysis} & CoNLL 2003~\cite{sang2003introduction} & Question & Tag sequence \\
                & \textit{Word Segmentation} & MSR~\cite{gao-etal-2005-chinese} & Question & Segmentation \\
                \midrule
              \multirow{2} * {\it Text Generation} & Semantic Parsing & WikiSQL~\cite{zhongSeq2SQL2017} & Question, Table & SQL \\
               & Summarization & CNN/Daily Mails~\cite{see-etal-2017-get} & Text & Text \\
			\bottomrule
		\end{tabular}
    \end{adjustbox}
	\caption{Some existing NLP datasets with their input forms and annotation forms.}
	\label{tab:study-existing-dataset-components}
\end{table*}

\begin{table*}
    \setlength{\dashlinedash}{0.5pt}
    \setlength\dashlinegap{1.5pt}
    \setlength{\dashlinegap}{4.5pt}
    \small
    \centering
    \begin{adjustbox}{width=\linewidth}
    \begin{tabular}{cp{5.4cm}p{5.4cm}c}    
    \toprule
     Component Type & How to Use                                                                                     & Corresponding Components in Tasks                                     & Example \\
    \midrule
    \multicolumn{4}{l}{\cellcolor{gray!40}{\textbf{Input Form}}}  \\ 
     Text           & Present questions                                                                                & Questions in QA or passage in Summarization. & \Cref{fig:interface-examples} (a)  \\ \hdashline
     Table          & Present tables                                                                                   & Tables in Text-to-SQL or TableQA task.      & \Cref{fig:interface-examples} (d) \\
    \midrule
    \multicolumn{4}{l}{\cellcolor{gray!40}{\textbf{Annotation Form}}} \\
    Text           & Type in the text                                                                                 & Summaries in Summarization or translations in Machine Translation.  & \Cref{fig:interface-examples} (a) \\ \hdashline
     Multi-Choice   & Select an option                                                                                 & Inference choices for NLI.                                  & \Cref{fig:interface-examples} (b)  \\ \hdashline
     Segmentation   & Use cursors to highlight text span                                                               & Answer span for Reading Comprehension.                & \Cref{fig:interface-examples} (c) \\ \hdashline
     Dropdown       & Select an option in the dropdown menu, then use cursors to highlight the corresponding text span & Entities and their pos-tags in POS Tagging.                          & \Cref{fig:interface-examples} (e) \\ \hdashline
     Slider         & Move the Slider                                                                                  & Scores in Textual Semantic Similarity.          & \Cref{fig:interface-examples} (f) \\
    \bottomrule
    \end{tabular}
    \end{adjustbox}
    \caption{Components provided in \sysname to support the construction of customized annotation interfaces.}
    \label{tab:interface-components}
\end{table*}


(2) \sysname boosts annotation efficiency with diverse options for annotation suggestion, as shown in \Cref{fig:concept-diagram}.
First, the flexibility of our annotation interface allows researchers to collect data for multiple sub-tasks at the same time. Correspondingly, \sysname includes a multi-task active learning (MTAL) system for annotation suggestions in 
such a setting. 
Hence, researchers can adopt a single versatile active learning model which reuses parameters, rather than having multiple models each dedicated to a single task.
Second, \sysname can collect certain annotator demographic features while collecting annotations. We construct another active learning system that leverages these demographic features, such as the age of the annotators, to make suggestions better aligned with the annotators' intentions.
Third, given the recent advances in large language models (LLMs) \citep{brown2020language}, our system also provides the ability to prompt LLMs using an API.


We conduct multiple experiments to demonstrate the effectiveness of our interface and the diverse annotation options it provides.
Additionally, through user studies from the perspective of both researchers and annotators, we demonstrate that \sysname is flexible and can meet the various needs of NLP researchers, as well as accelerate the annotation process for annotators. The video link to our demo system is at \url{https://www.youtube.com/watch?v=0trF6zpIax0}. 

\section{Front-End}
\label{sec: front-end}
Many tasks in NLP can be decomposed into a small set of basic actions.
For example, given a piece of text, an annotator should select from a fixed set of labels (e.g. NLI) or provide a numeric score (e.g. sentiment analysis).
We surveyed several existing NLP datasets (in \Cref{tab:study-existing-dataset-components}) to identify common basic actions.
We then developed several modular front-end components (in \Cref{tab:interface-components}) which researchers can use to construct custom interfaces for presenting examples to annotate and collecting annotations.
These components cover many comprehensive NLP annotation tasks: \Cref{fig:interface-examples} in \Cref{appendix: more-details-about-our-system} shows several interfaces for common NLP tasks constructed using these components.
These interfaces are also provided in \sysname.

\begin{figure}[t]
    \centering
    \includegraphics[width=\linewidth]{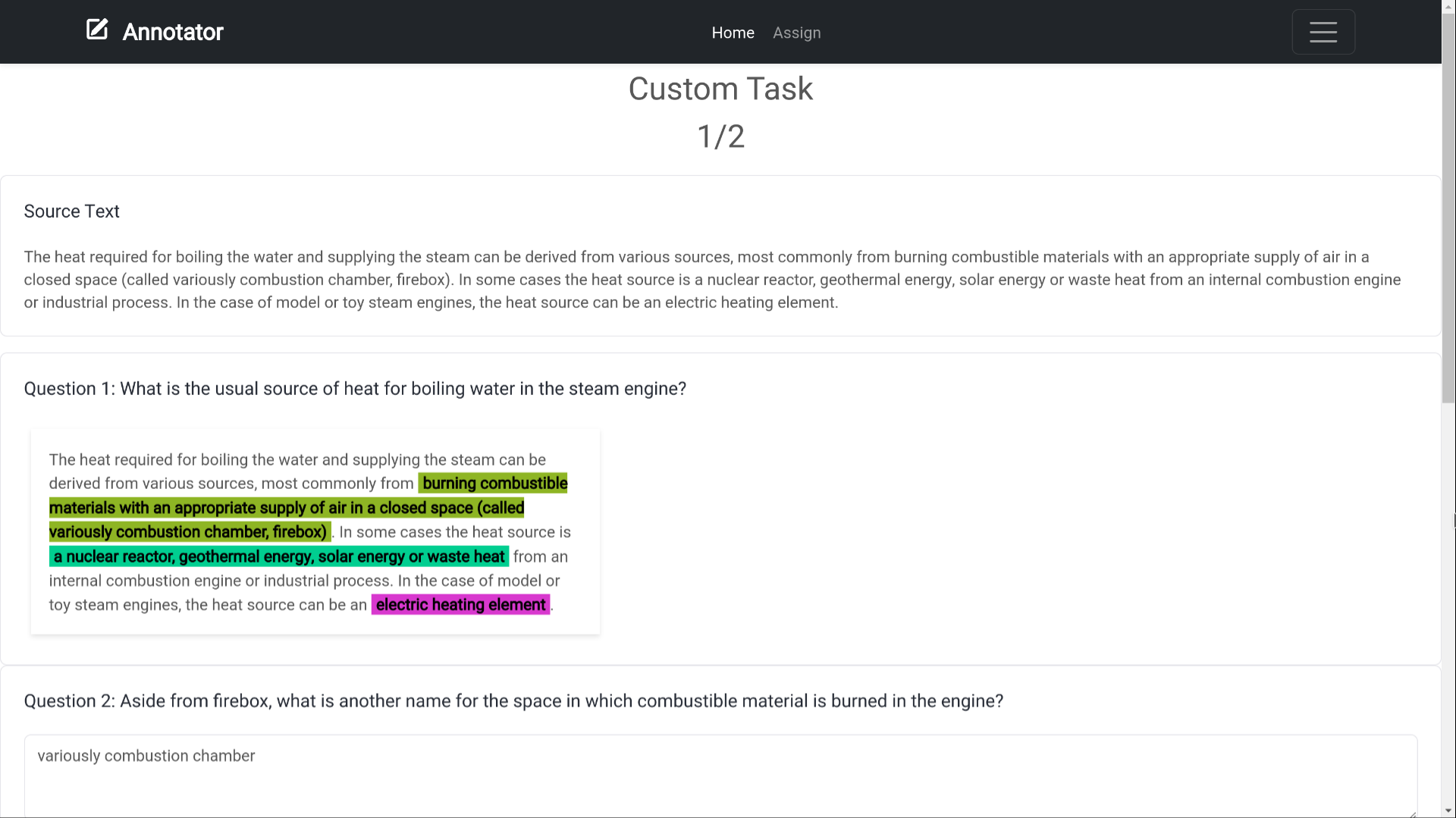}
    \caption{A custom \sysname annotation interface consisting of the Segmentation and Text components. This interface requires the annotator to complete two tasks, evidence selection and free-form question answering, at the same time for one instance.}
    \label{fig:customized}
\end{figure}

\paragraph{Building Custom Interfaces}
In addition, researchers can build their own custom interfaces by simply specifying the desired components in a JSON file.
These components can be added in any order or repeatedly to meet researchers' needs.
\Cref{fig:customized} provides a simple example of a custom interface, consisting of the Segmentation and Text components.
The corresponding JSON to construct this interface is presented in \Cref{appendix: customized-json}.
\Cref{appendix: more-details-about-our-system} presents more details about annotation procedures and our implementation.

\paragraph{Survey of Front-End Effectiveness}
To further motivate and validate the effectiveness of our modular front-end components and custom interface building, we surveyed ten NLP researchers, asking them to design a project where they need to collect data, and to describe in detail the types of data they want to collect.
The responses from this survey (in \Cref{appendix: front-end-user-study}) demonstrate that \sysname can fulfill researchers' needs on a variety of tasks.

\section{Back-End}
\label{sec: back-end-system}

To match the flexibility of the front-end interface, we provide several back-end options to assist the annotator. In the Multi-Task Active Learning back-end (\Cref{sec: mtal-backend}), annotators annotate sub-tasks at the same time for the same instance, and we train a single back-end model on those sub-tasks simultaneously.
The Active Learning with Demographic Features back-end (\Cref{sec: al-demographics-backend}) provides another option for active learning, which leverages demographic information collected from the annotator to improve the accuracy of annotation suggestions.
With the surge of large language models (LLMs), we also introduce a Prompt Query System as an alternative back-end (\Cref{sec: prompt-backend}), which queries the API of certain LLMs to provide annotation suggestions.

\section{Back-End Option 1: Multi-Task Active Learning}
\label{sec: mtal-backend}

Some annotation tasks in NLP may require annotating multiple types of information for the same input, e.g. providing part of speech tags and labeling named entities in a sentence.
In the process of providing answers to multiple sub-tasks for the same input, annotators may be able to annotate these sub-tasks more quickly when done together on the same input, rather than separately, due to their growing familiarity with the sub-tasks through the annotation process.
To test this hypothesis, we conduct a user study with six annotators, who were tasked with annotating chunk parsing and named entity recognition for 20 sentences from the CoNLL2003 shared task \citep{tjong-kim-sang-de-meulder-2003-introduction}.
We measured the time it took for them to annotate the two sub-tasks separately versus together for the same input.

\begin{figure}[t]
    \centering
    \includegraphics[width=0.9\linewidth]{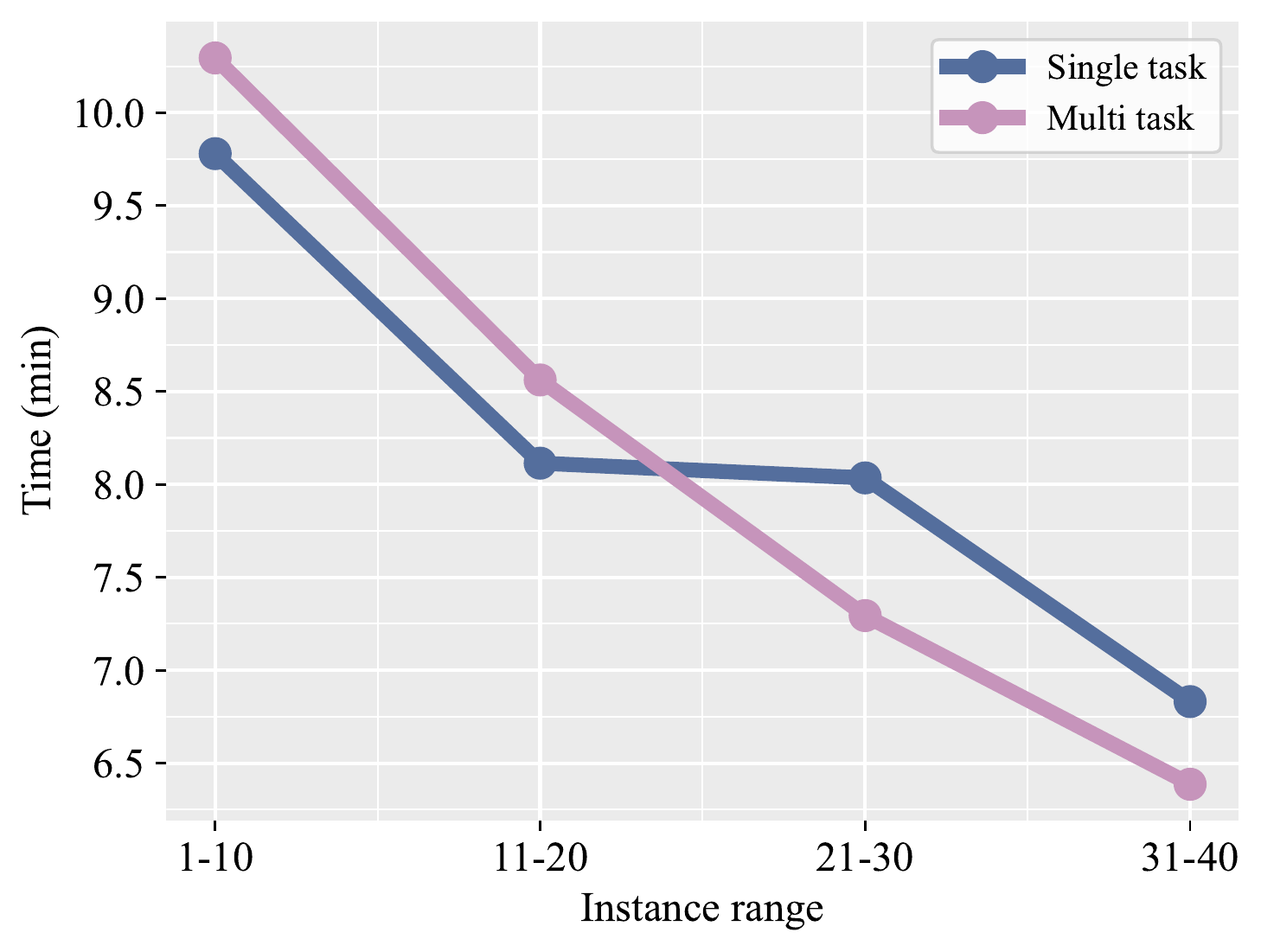}
    \caption{Average annotation time for user studies on CONLL2003 datasets. For ``Single task'', annotators annotate chunk parsing and named entity recognition sub-tasks separately, and we calculate the sum of their annotation time. For ``Multi task'', annotators annotate the two sub-tasks together for the same instance.
    The time for each range of 10 annotations is the sum of time for 5 chunk parsing tasks and 5 NER tasks.
    }
    \label{fig:user-study-time}
\end{figure}

We find that annotators tend to spend less time as they become more familiar with the tasks, as shown in \Cref{fig:user-study-time}.
In addition, annotators become more efficient when annotating the sub-tasks together instead of annotating them individually once the annotator gets more familiar with the task. \Cref{appendix: user-studies-result} provides more details of the study.

Leveraging this result, we build a multi-task active learning back-end system to assist annotators on the front-end.
We employ a single versatile model that can suggest annotations for various tasks, which has benefits over training multiple models each dedicated to an individual task.

\subsection{Method}

Given a model with a linear prediction layer, we define the probability of the model generating an answer to be the softmax of logits from the linear layer.
That is, given the logits $\{h_1, h_2, \cdots, h_n\}$, where $n$ is the number of possible answers, the probability of selecting answer $i$ is:
$$
    P_i = \frac{e^{h_i}}{\sum_{j=1}^n e^{h_j}}
$$
Following the uncertainty sampling method from \citet{lewis1995sequential, settles2009active}, we query the instance whose best labeling (the predicted label in our case) is the least confident.
The flexibility of the annotation interface builder allows for collecting data for different tasks for the same input at the same time. Therefore, we calculate a joint loss $L$ over multiple tasks:

$$
    L = \sum_{i=1}^m\alpha_iL_i,
$$
where $L_i$ is the loss for the $i$th task, $\alpha_i$ is a weight hyperparameter.

\subsection{Experiments}
\label{subsec: mtal-vs-al}


We conduct experiments to investigate the performance of single-task vs. multi-task active learning using \sysname.
The data for these experiments are from the CoNLL 2003 shared task \citep{tjong-kim-sang-de-meulder-2003-introduction}, which includes chunk parsing and named entity recognition. \Cref{appendix: mlal-set-ups} provides more details about the models we use for our experiments.
We train a single model jointly on both tasks and compare with models trained separately on each task.



\paragraph{Results}
\Cref{tab: multi-task-al-conll2003} shows models' performance on the original CoNLL 2003 test set. 
Active learning on each task separately yields slightly better results on named entity recognition, while multi-task active learning with \sysname performs better on chunk parsing.
In terms of performance, the difference is only about or within 1 percent.
However, we see substantial gains in terms of training time: \sysname trains in around half of the time than training on each task separately.
Thus, multi-task active learning from \sysname saves GPU hours while still performing comparably to active learning on each task separately.
In \Cref{appendix: conll2003-results},~\Cref{fig:conll2003-acc,fig:conll2003-recall,fig:conll2003-precision,fig:conll2003-f1} show the model performance on the validation set as the number of instances increases. \sysname can achieve an accuracy of 95\% and 80\% on chunk parsing and named entity recognition after tuning on 1,000 examples, respectively. After this point, \sysname can start suggesting annotations with high accuracy and f1 scores.




\begin{table}[t]
    \centering
    \setlength{\dashlinedash}{0.5pt}
    \setlength{\dashlinegap}{4.5pt}
\scalebox{0.75}{
    \begin{tabular}{llccc}
    \toprule
                         Setting  &   Task   & Acc     & F1   & Time (h)        \\
    \midrule
    \multirow{2}{*}{AL} & CP           & 94.07 \tiny{0.07} & 93.57 \tiny{0.07} & \multirow{2}{*}{7.61 \tiny{0.16}}\\
    & NER          & \textbf{97.70 \tiny{0.04}}  & \textbf{90.18 \tiny{0.23}} & \\
    \hdashline
    \multirow{2}{*}{\sysname$_{\text{o1}}$} & CP& \textbf{94.20 \tiny{0.07}}  & \textbf{93.73 \tiny{0.08}} & \multirow{2}{*}{\textbf{4.28 \tiny{0.03}}}\\
    & NER & 97.33 \tiny{0.23} & 88.77 \tiny{0.78} & \\
    \bottomrule
    \end{tabular}
    }
    \caption{Model performance on chunk parsing (``CP'') and named entity recognition (``NER''). ``AL'' refers to active learning, \sysname$_{\text{o1}}$ refers to the first option of multi-task active learning in \Cref{sec: mtal-backend}. ``Acc'', ``F1'' refer to accuracy and f1 scores, respectively. The experiments are on a single NVIDIA Tesla V100 GPU. We run each experiment 3 times and report the average scores and the standard deviation as the footnote.}
    \label{tab: multi-task-al-conll2003}
\end{table}

\section{Back End Option 2: Active Learning With Demographic Features}
\label{sec: al-demographics-backend}

Prior work has shown that annotators may have different opinions in tasks such as sentiment analysis or natural language inference \citep{plank-etal-2014-linguistically, aroyo2015truth, pavlick2019inherent}. In those cases, annotators' demographic information has been shown to be correlated with their annotations \citep{al-kuwatly-etal-2020-identifying}. Thus, we built \sysname's customizable interface with the ability to collect basic demographic features of the annotators, such as age, race, or gender. On the back-end, we develop an active learning method that incorporates annotators' demographic features to make personalized annotation suggestions.

\subsection{Method}
Following \citet{geva-etal-2019-modeling}, we concatenate the demographic features $(d_1, d_2, \cdots, d_m)$ to the sentence tokens $(w_1, w_2, \cdots, w_n)$ and feed the concatenated sequence $(d_1, d_2, \cdots, d_m, w_1, w_2, \cdots, w_n)$ to the model to make a prediction, where $d_j$ is a single value for one type of demographic feature (e.g. age), $w_i$ is a token in the input instance. We follow the same active learning paradigm as in multi-task active learning (\Cref{sec: mtal-backend}).

\subsection{Experiments}
We conduct experiments to investigate the performance of active learning when augmenting an input passage with the annotators' age as a demographic feature.
We use the sentiment analysis dataset of \citep{diaz2018addressing}, a 5-way classification dataset. \Cref{appendix: demographic-experiment-set-ups} gives more details about the datasets and our model.


\paragraph{Results} 
There has been a recent trend in our community to collect different annotations from different annotators on the same instance \cite{plank-2022-problem}. Our experiments show that under such a setting, providing annotation suggestions is difficult. If we only provide the input sentence to the active learning model, the model performs poorly, close to random performance (\Cref{tab: demographics}), as it might get confused with different labels given by different annotators on the same instance. Moreover, datasets such as the Sentiment Analysis dataset involve over 7k annotators; it is unrealistic to train an annotation suggestion model for each annotator. However, our active learning model performs substantially better when the demographic feature of annotator age is fed along with the input sentence (\Cref{tab: demographics}). Therefore, by leveraging some non-essential demographic features such as age, \sysname is able to make annotation suggestions that better align with annotators' intents. 

\begin{table}[t]
    \small
    \centering
    \begin{tabular}{ll}
       \toprule
       & \multicolumn{1}{c}{Acc} \\
       \midrule
    Random &  26.80 \tiny{0.86} \\
    Statement-Only & 29.61 \tiny{17.69} \\
    \sysname$_{\text{o2}}$ & \textbf{38.04 \tiny{8.08}} \\
    \bottomrule
    \end{tabular}
    \caption{Model performance on the Sentiment Analysis dataset \citep{diaz2018addressing} (5-way classification). For ``Random'', the model randomly chooses from the 5 labels; The inputs are: ``Statement-Only'': only the statement; 
    In \sysname$_{\text{o2}}$, we concatenate the statement with age as the demographic feature as discussed in \Cref{sec: al-demographics-backend}.}
    \label{tab: demographics}
\end{table}

\section{Back End Option 3: Prompt Query System}
\label{sec: prompt-backend}

In recent years, LLMs have demonstrated an undeniable capability to perform few-shot tasks and data augmentation \cite{brown2020language, yoo2021gpt3mix,huang2022large}. To this end, \sysname supports a third back-end option: leveraging LLMs to suggest annotations by querying via API calls.


\subsection{Method}
We devise a few-shot prompt consisting of example training annotations, followed by an instance where the model must provide the annotations.
The prompt structure we use is:

{
\small
\texttt{Given the sentence [Sentence$_1$] the [Task Name] are [Answer$_1$]\textbackslash n\textbackslash n Given the sentence [Sentence$_2$] the [Task Name] are [Answer$_2$]\textbackslash n\textbackslash n $\cdots$ Given the sentence [Sentence$_2$] the [Task Name] are},
}

\noindent An example prompt is shown in \Cref{appendix: prompt-examples}.
\sysname allows for specifying the number of few-shot examples as well as multiple methods for selecting these examples.
One way of constructing the prompt is by choosing examples randomly. Another way is to select similar examples from the training data to construct the prompt~\cite{Selective_Annotation}. Specifically, to select similar examples, we provide the input sentence from the training set and test set to the BERT model. We then take the $N$ training examples that have the highest cosine similarity scores to the example in the test set.

\subsection{Experiments}
\label{subsec: prompt-based-system-experiment}

To determine the effectiveness of these prompts, we conduct experiments where we vary the number of few-shot examples (5 or 10) and the methods for selecting examples for the prompt (random or similar).
In these experiments, we again use the CoNLL 2003 shared task data, and we query the API of the GPT3 \citep{brown2020language} text-curie-001 model (6.7B).


\paragraph{Results} Performance of the LLMs with various prompt settings is shown in \Cref{tab: prompt-conll2003}.
First, we find that the performance of LLMs on the POS tagging and NER tasks is much worse than the performance of a tuned active learning model (\Cref{tab: multi-task-al-conll2003}). This aligns with other observations that prompt-based methods are not appropriate for sequence labeling tasks \citep{hou2022inverse}.
The reason could be that sequence labeling tasks differ from text completion, and the LLMs lack the knowledge of the syntactic structure of the sentence in order to make correct predictions.
In addition, accuracy on named entity recognition is unproportionally high, because the model ``cheats'' by predicting many ``O''s (in BIO tagging) instead of the actual entities. 
However, due to the input length limit to the API of GPT3, we cannot feed an unlimited number of examples, which limits the power of utilizing prompts for annotation suggestions. Our results show that the prompt might not be suitable to be used as an annotation suggestion for tasks such as sequence labeling, but researchers might find prompts useful in other tasks, especially text generation \citep{chen-etal-2022-coherent, chen2022unisumm}.

\begin{table}[t]
\small
\centering
\begin{tabular}{llrrrr}
   & Setting & \multicolumn{1}{c}{Acc} & \multicolumn{1}{c}{R} & \multicolumn{1}{c}{P} & \multicolumn{1}{c}{F1}    \\
   \toprule
\multirow{4} * {\textit{CP}} & Random$_5$  & 18.58    & 18.37  & 5.26      & 8.17  \\
&Random$_{10}$ & \textbf{31.34}    & 25.84  & \textbf{6.71}      & \textbf{10.66} \\
&Sim$_{5}$ & 24.45 & 26.07 & 4.62 & 7.85 \\
&Sim$_{10}$ & 25.54 & \textbf{26.22} & 4.48 & 7.65 \\
\midrule
\multirow{4} * {\textit{NER}} &Random$_5$  &  55.24     & 1.10   & 1.49      & 1.27   \\
&Random$_{10}$ & \textbf{64.66}    & 4.41   & \textbf{6.24}      & \textbf{5.17} \\
&Sim$_{5}$ & 59.26 & \textbf{8.85} & 0.49 & 0.92\\
&Sim$_{10}$ & 53.09 & 8.68 & 0.52 & 0.97\\
\bottomrule
\end{tabular}
\caption{Prompt performance on chunk parsing (CP) and named entity recognition (NER) for settings of random selection (Random) and selecting the most similar examples (Sim) from the training set.  Footnote $_5$ and $_{10}$ denote the number of examples selected to construct the prompt query. 
}.
\label{tab: prompt-conll2003}
\end{table}

\section{Related Work}

As \sysname is designed to be accessible, flexible, and efficient for annotators, we briefly compare our tool with existing annotation tools in these aspects.
First, regarding accessibility, popular existing paid annotation tools such as Prodigy\footnote{\url{prodi.gy}} are hundreds of dollars for a single license, which may be cost-prohibitive for small research labs or hobbyists.
In contrast, our tool is free and open-source, which is ideal for researchers to use and extend for their own annotation purposes.

Second, regarding flexibility, many existing tools have been developed for individual tasks, such as coreference resolution \citep{bornstein-etal-2020-corefi} or sentiment analysis \citep{yimam-etal-2020-exploring}.
Recently, there has been work developing tools that support multiple tasks within the same framework, including Doccano \citep{nakayama2018doccano} and Crowdaq \citep{ning-etal-2020-easy}.
Others have extended this concept to allow the annotation of multiple tasks at the same time, including WebAnno \citep{yimam-etal-2013-webanno} and LexiClean \citep{bikaun-etal-2021-lexiclean}.
Similar to these tools, \sysname provides several modular front-end components which researchers can use to construct annotation interfaces for their own needs.
Furthermore, \sysname does not need technical knowledge, as custom interfaces can simply be defined in JSON.

One of \sysname's main contributions is its support of multi-task active learning, which is not supported by the above tools.
For an overview of active learning, we refer the reader to the survey of \citet{settles2009active}. 
Recent work has focused on active learning in annotation tools to save annotators' time and effort.
These include INCEpTION \citep{klie-etal-2018-inception}, which used uncertainty sampling, and DART \citep{chang-etal-2020-dart}, which proposed a retrieval-based method to suggest labels to the annotator.
\citet{ein-dor-etal-2020-active} also presents an analysis of different example selection methods under data imbalance scenarios.
\sysname supports \textit{multi-task} active learning, where a model can learn from different task annotations of the same sentence or document.
Most comparable to our tool are ActiveAnno \citep{wiechmann-etal-2021-activeanno} and POTATO \citep{pei-etal-2022-potato}, which also support modular components and multi-task active learning.
However, these papers do not present experiments validating their multi-task active learning setup.
In contrast, we perform extensive experiments on multi-task active learning with human annotators.







\section{Conclusion}

We present our annotation system \sysname, which is designed to be flexible, customizable, and support efficient annotation collection. \sysname consists of a customizable front-end interface and multiple back-ends for suggesting annotations. Through user studies, we demonstrate that our modular front-end components can be used to construct custom interfaces that support various needs of data collection by NLP researchers. We also conducted several experiments that showcase \sysname's backends for multi-task active learning, demographic feature based active learning, and LLM prompting.
Our work continues a recent trend of building customizable front-end annotation interfaces.
Moreover, we design efficient back-end systems that can take advantage of the custom interface as well as the recent advances of LLMs to more efficiently gather annotations. Future work might further consider how to leverage the power of LLMs to ease the annotation process for annotators.



\section{Limitations}

We demonstrated that the front-end components we provide in \sysname can cover a wide range of data collection. But further investigation might be needed to study annotation for certain custom tasks, especially tasks involving languages other than Chinese and English. We welcome additional contributions to \sysname.

Due to budget, we only tested the LLM prompts with \sysname using the GPT3 text-curie-001 model, but our framework can be easily adapted to query other LLMs. With advances in LLMs, researchers might get better performance on tasks such as sequence labeling by using more powerful models such as ChatGPT.

\section{Ethics}



\paragraph{Large Language Models (LLMs)} 
We use one GPT-3.5 model, \texttt{text-curies-001}, to provide annotation options in \sysname.
We only use \texttt{text-curies-001} for experiments and analysis, which are presented in \S~\ref{sec: prompt-backend}.
Other usage of GPT-3 or its family (e.g., ChatGPT) is not included.

\paragraph{Human Evaluation} We surveyed ten NLP researchers, who are currently holding research positions in NLP and have published articles in *ACL venues, to demonstrate that \sysname can cover diverse NLP tasks.
The annotation study is conducted by six college students voluntarily, who are to complete or have completed courses from universities in the U.S.


\bibliography{anthology,custom}
\bibliographystyle{acl_natbib}

\newpage

\appendix

\section{More Details about Our System}
\label{appendix: more-details-about-our-system}

\begin{figure}[t]
    \centering
    \includegraphics[width=\linewidth]{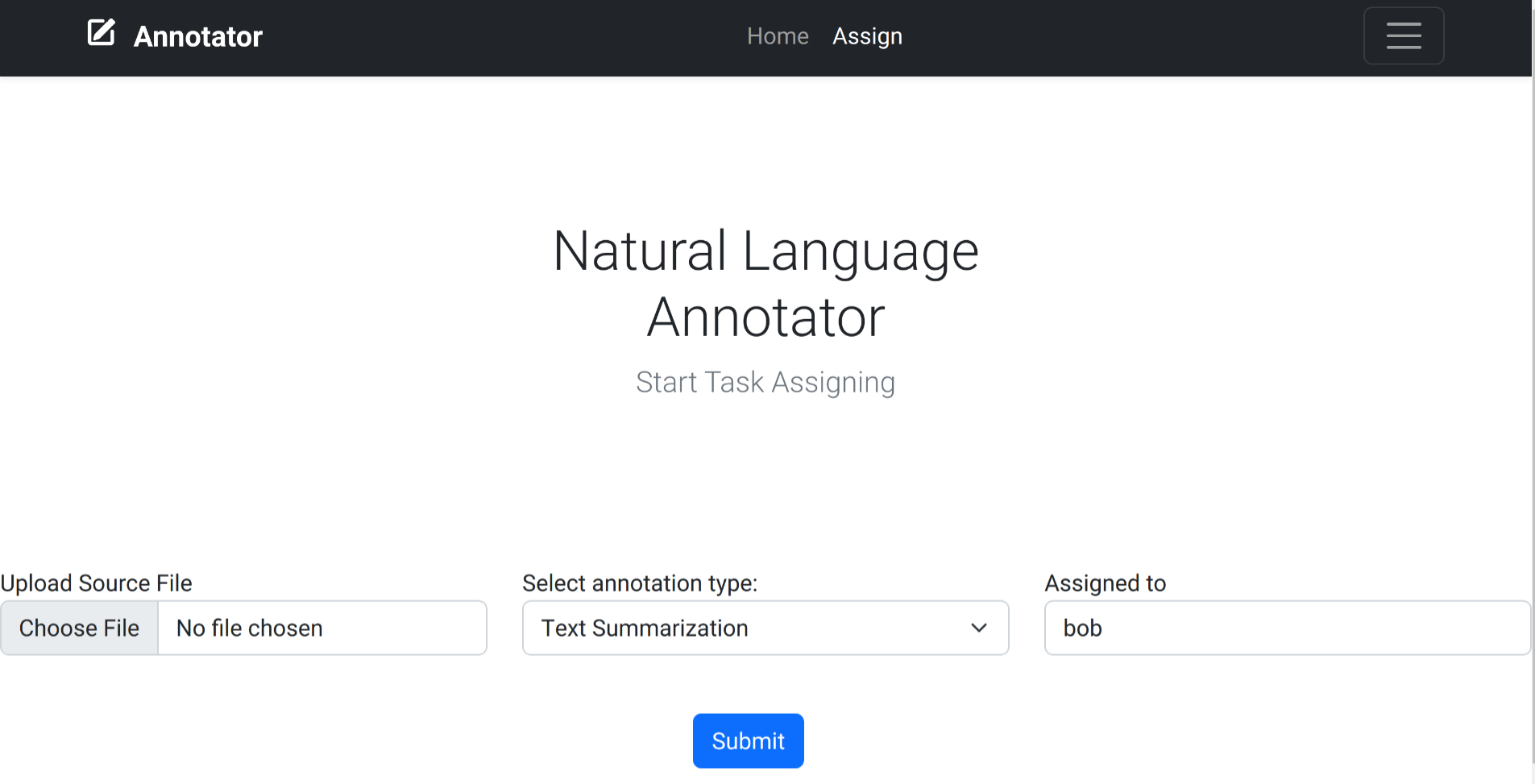}
    \caption{The administrator uploads the data to be annotated, selects the ``annotation type'' by choosing pre-defined (``Text Summarization'' in this figure) or his customized version of the interface, and assigns to the annotator ``bob''. }
    \label{fig:task-assignment}
\end{figure}

\begin{figure*}[t]
    \centering
    \subfigure[]{\includegraphics[height=3.7cm]{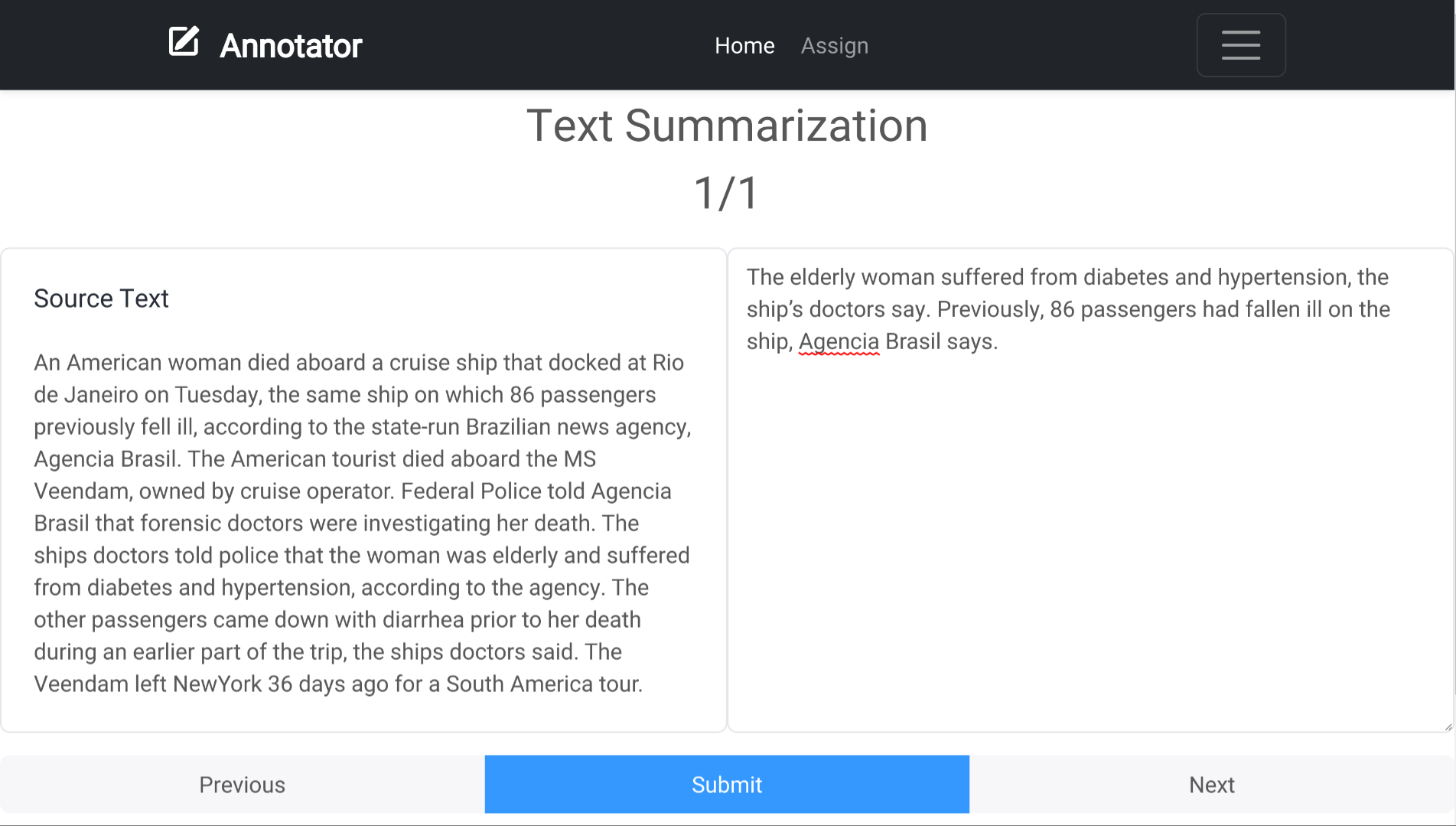}} 
    \subfigure[]{\includegraphics[height=3.7cm]{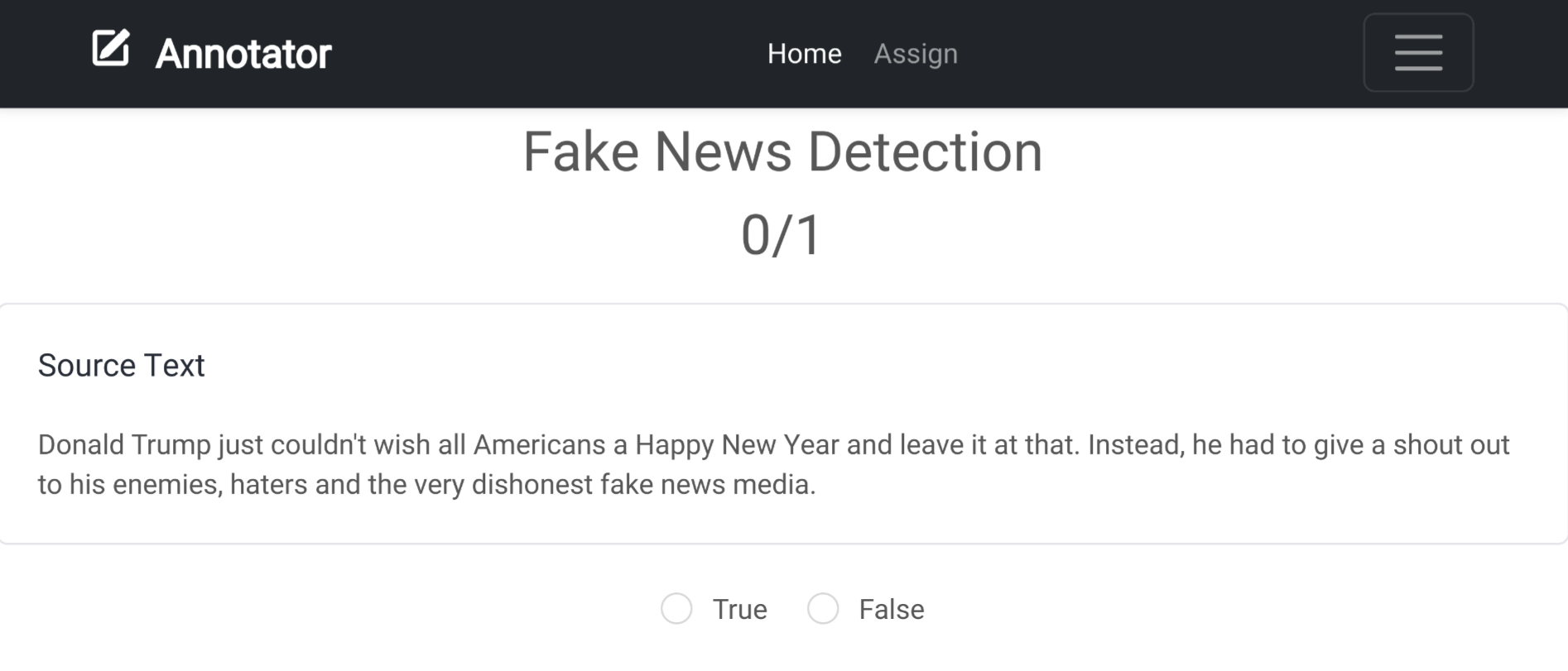}}
    \subfigure[]{\includegraphics[height=3.7cm]{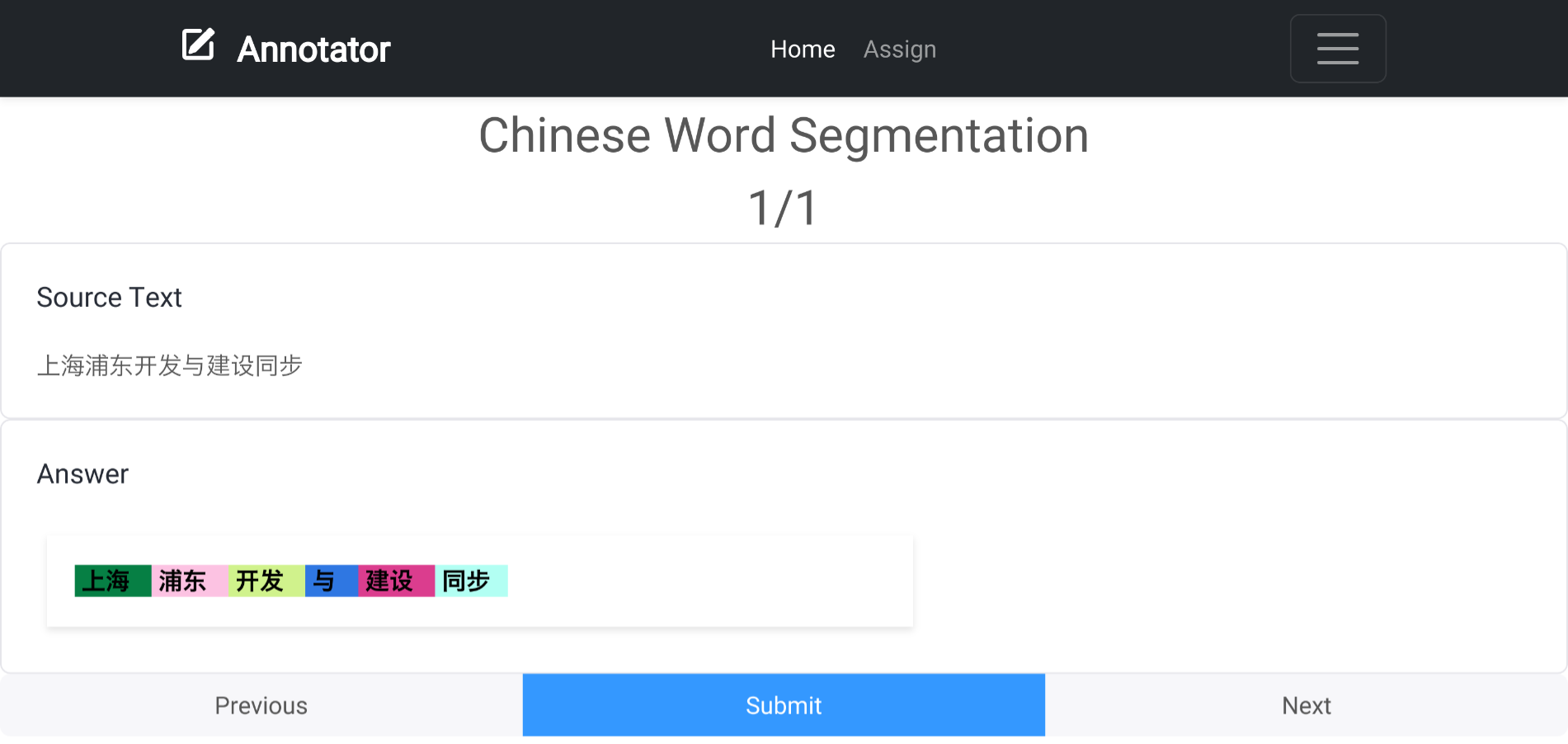}} 
    \subfigure[]{\includegraphics[height=3.7cm]{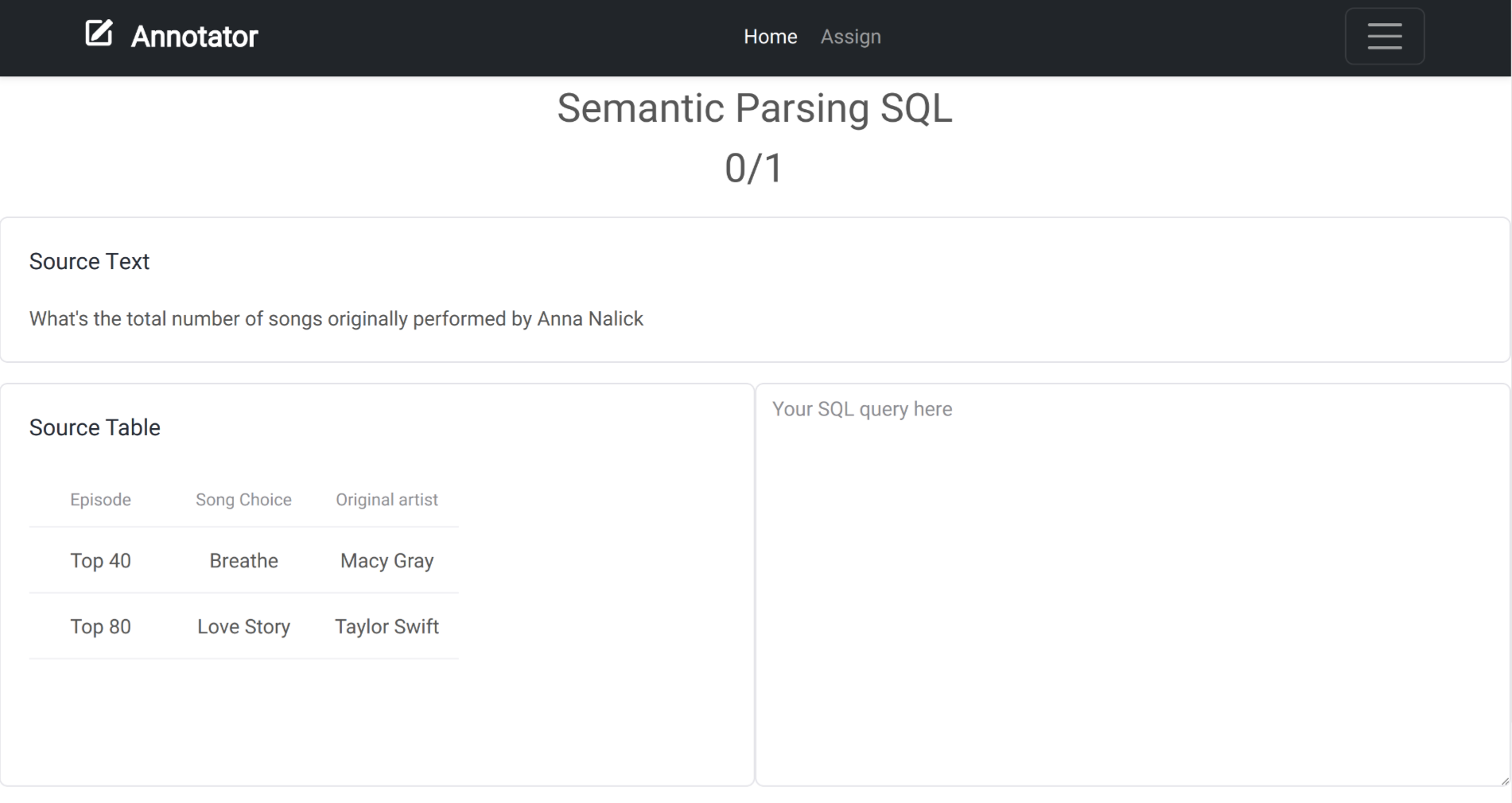}} 
    \subfigure[]{\includegraphics[height=4cm]{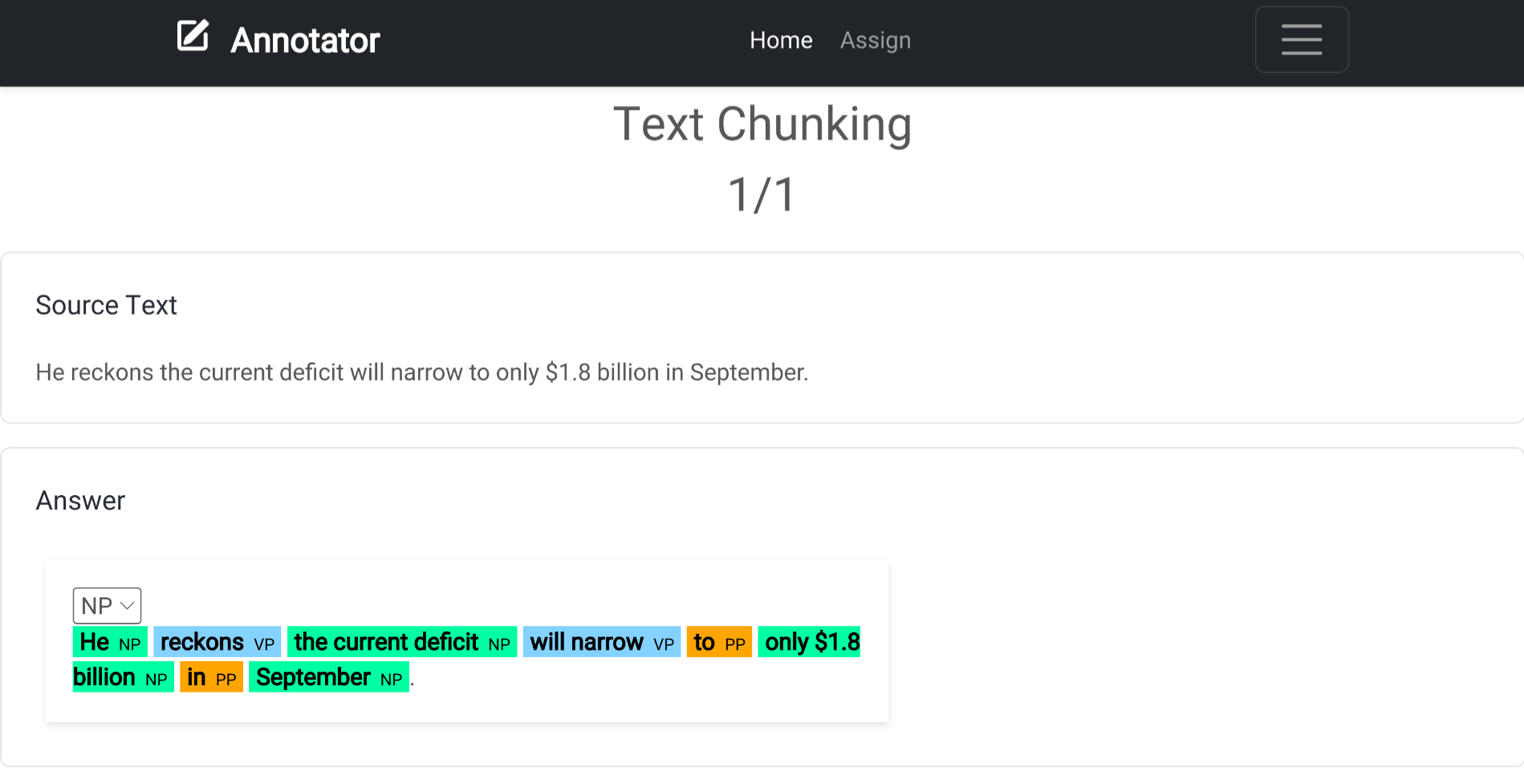}} 
    \subfigure[]{\includegraphics[height=4cm]{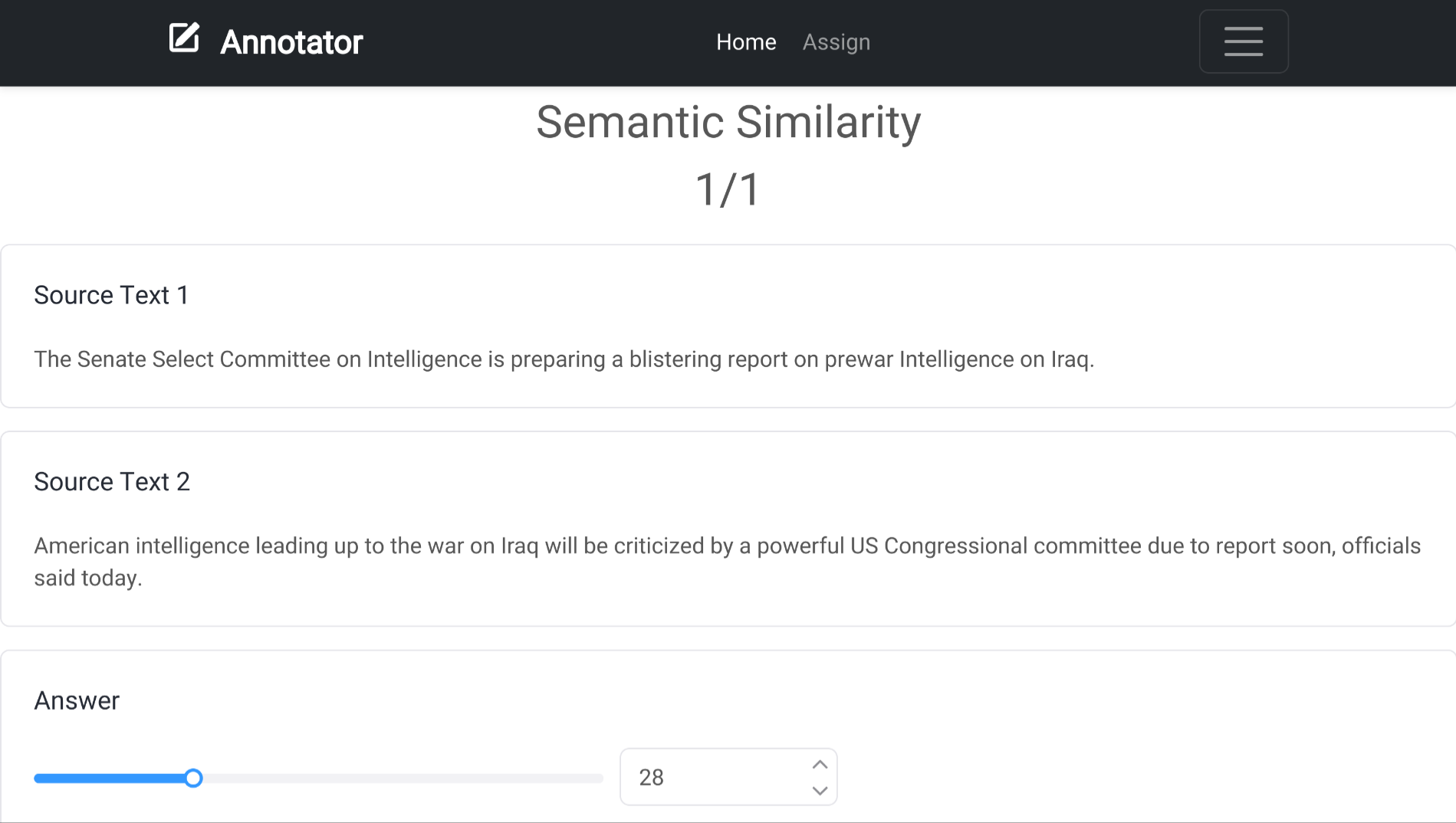}} 
    \caption{Annotation interfaces constructed by \sysname for (a) text summarization; (b) fake news detection; (c) Chinese word segmentation; (d) text-to-SQL; (e) text chunking; (f) semantic similarity.}
    \label{fig:interface-examples}
\end{figure*}

\paragraph{Annotation Interface Examples}
\Cref{fig:interface-examples} shows some basic annotation interface examples constructed by \sysname, where the basic components correspond to those described in \Cref{tab:interface-components}.

\paragraph{Annotation Procedure}
We define two types of roles: administrator and annotator in the system. The administrator is the person who composes the annotation interface, while annotators are the people who annotate the data.
A typical annotation process is:
\begin{enumerate}[leftmargin=0.5cm]
\item The administrator designs the task format and uploads the dataset to be annotated.
\item The administrator creates and assigns tasks to specific users, as shown in \Cref{fig:task-assignment}.
\item The annotators annotate the data with annotations suggested by our back-end system (discussed in \Cref{sec: back-end-system}).
\item The administrator exports the annotated data in JSON format. 
\end{enumerate}

\paragraph{Implementation and Logistics}
The annotation interface is built using the React.js framework and various open-source web components. The interface is responsive and can adapt to many screen sizes. The server-side back-end system uses the Flask framework in Python to serve REST APIs for interaction with the front-end interface. We use a SQLite database to store and manage annotation data. For tasks with active learning support, we define a REST API interface for interaction with the back-end active learning model using sockets, threads, and JSON (\Cref{appendix: customized-json} shows an example JSON file for the customized interface). With this new REST API interface, new models can be connected easily.

\section{Multi-Task User Studies}
\label{appendix: user-studies-result}

\begin{figure}[t]
    \centering
    \includegraphics[width=0.9\linewidth]{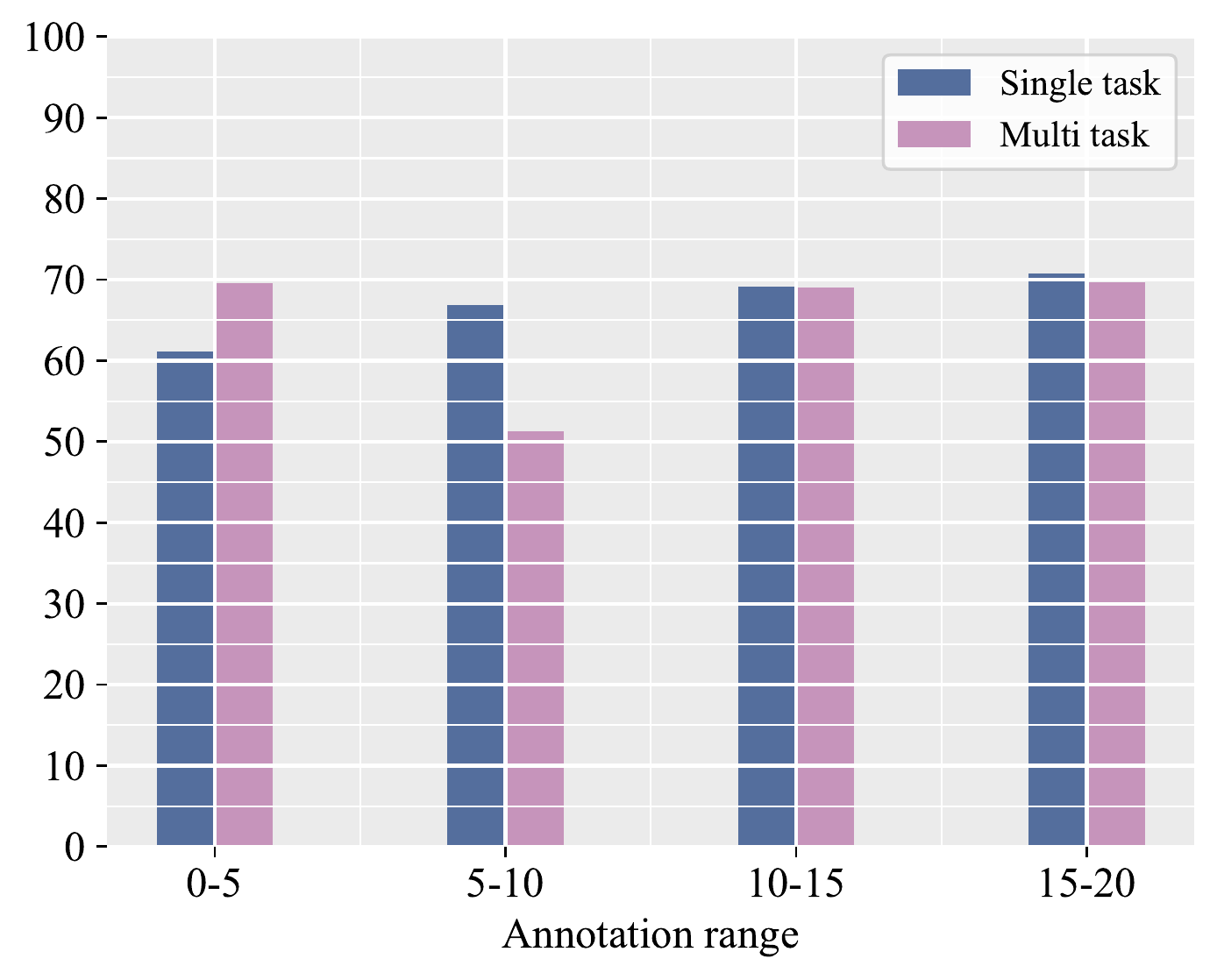}
    \caption{Accuracy scores of annotators' performance on 30 chunk parsing examples. ``Single task'' corresponds to the time they spend on the examples when they annotate the 2 tasks separately, while ``Multi task'' corresponds to the time they spend when they annotate the two tasks (chunk parsing and named entity recognition) for the same instance at the same time
    .}
    \label{fig:user-study-pos-acc}
\end{figure}

\begin{figure}[t]
    \centering
    \includegraphics[width=0.9\linewidth]{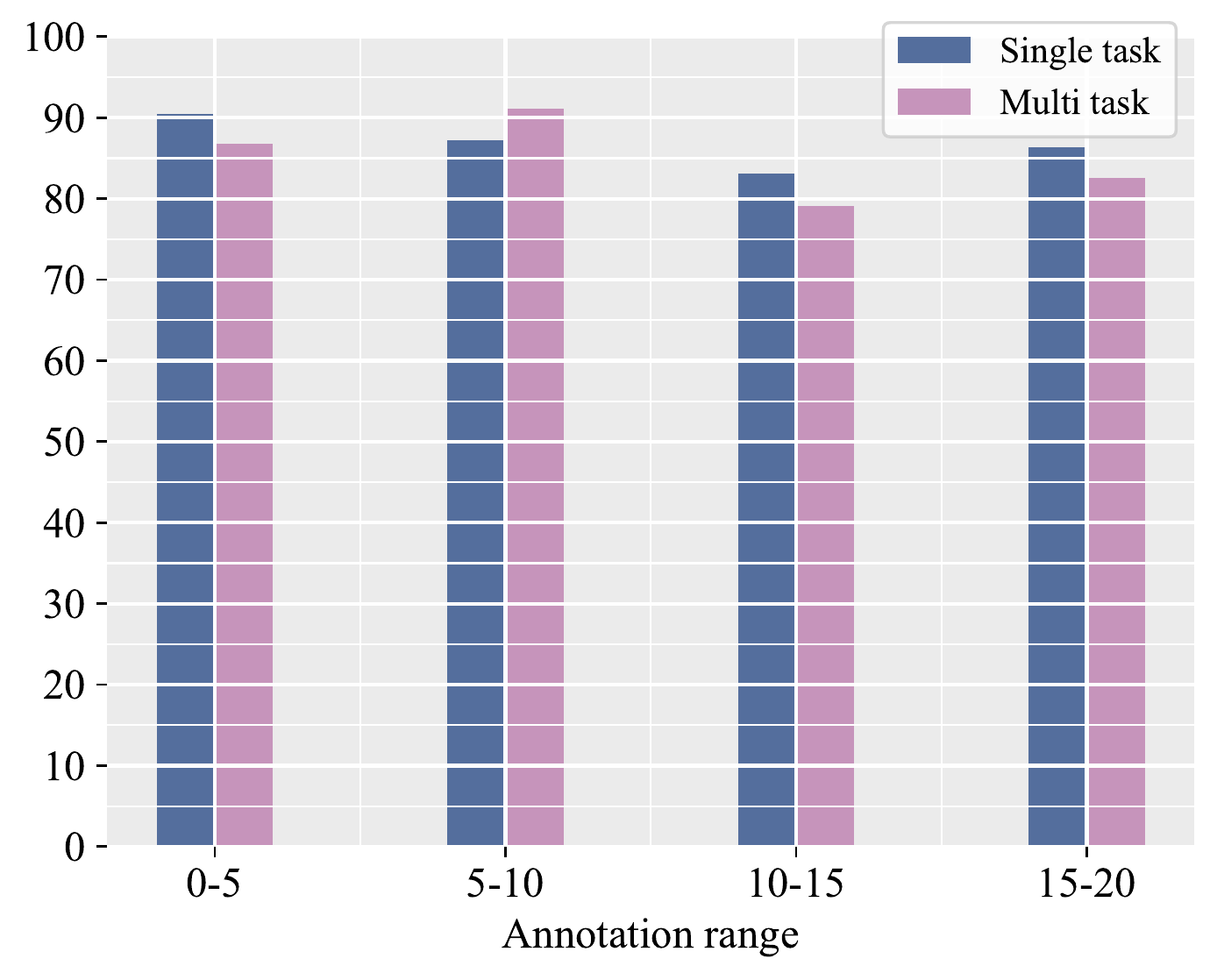}
    \caption{Accuracy scores of annotators' performance on 30 named entity recognition examples.}
    \label{fig:user-study-ent-acc}
\end{figure}

\begin{figure}[t]
    \centering
    \includegraphics[width=0.9\linewidth]{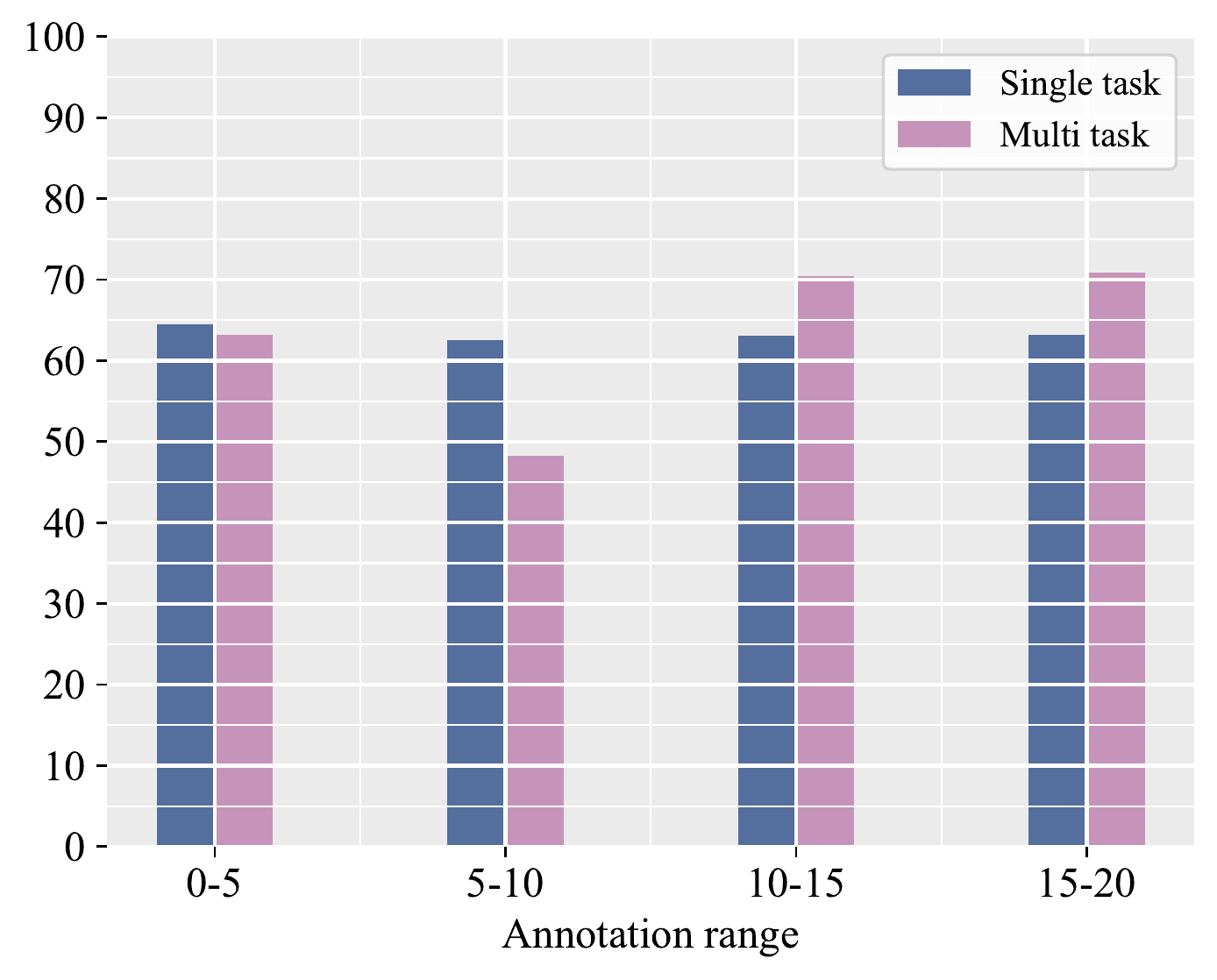}
    \caption{F1 scores of annotators' performance on 30 chunk parsing for user studies.}
    \label{fig:user-study-pos-f1}
\end{figure}

\begin{figure}[t]
    \centering
    \includegraphics[width=0.9\linewidth]{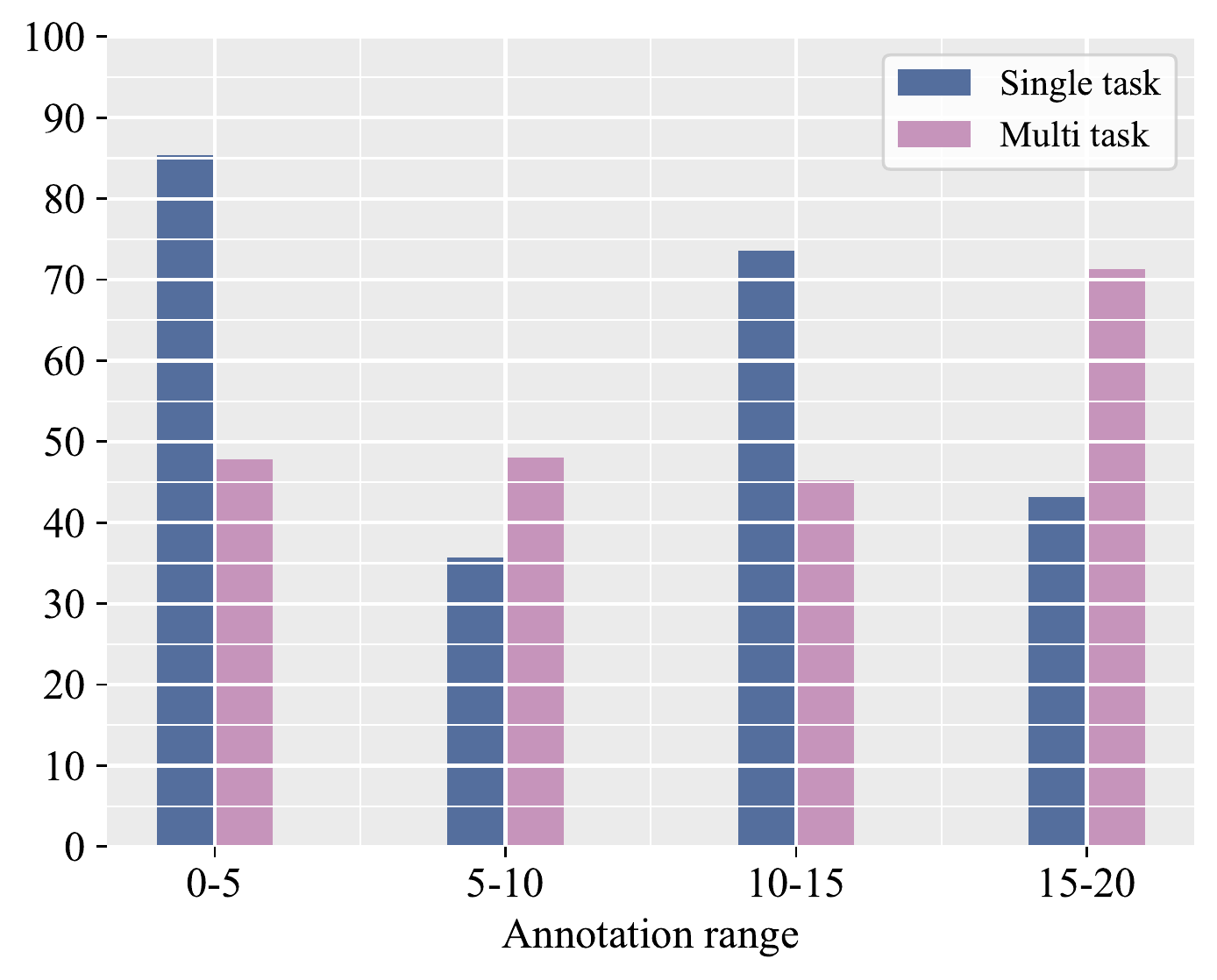}
    \caption{F1 scores of annotators' performance on 30 named entity recognition examples.}
    \label{fig:user-study-ent-f1}
\end{figure}

\Cref{fig:user-study-ent-acc,fig:user-study-pos-acc,fig:user-study-ent-f1,fig:user-study-pos-f1} show the annotators' performance on the tasks of chunk parsing and named entity recognition. We compare the annotations we collect to the ground truth labels from the original CoNLL2003 to calculate the annotation performance.
In general, annotators perform about the same when annotating the two tasks separately v.s. annotating tasks for the same instance at the same time in terms of accuracy and f1 scores. Note that there is a big fluctuation in the f1 score for named entity recognition because of the way we calculate the score. The f1 score only considers tags that are not ``O'' into account, small variations in annotations might lead to a big f1 score difference then.

\section{Experiment Set-ups for Back-End Option 1: Multi-Task Active Learning}
\label{appendix: mlal-set-ups}

\begin{figure}[t]
    \centering
    \includegraphics[width=0.9\linewidth]{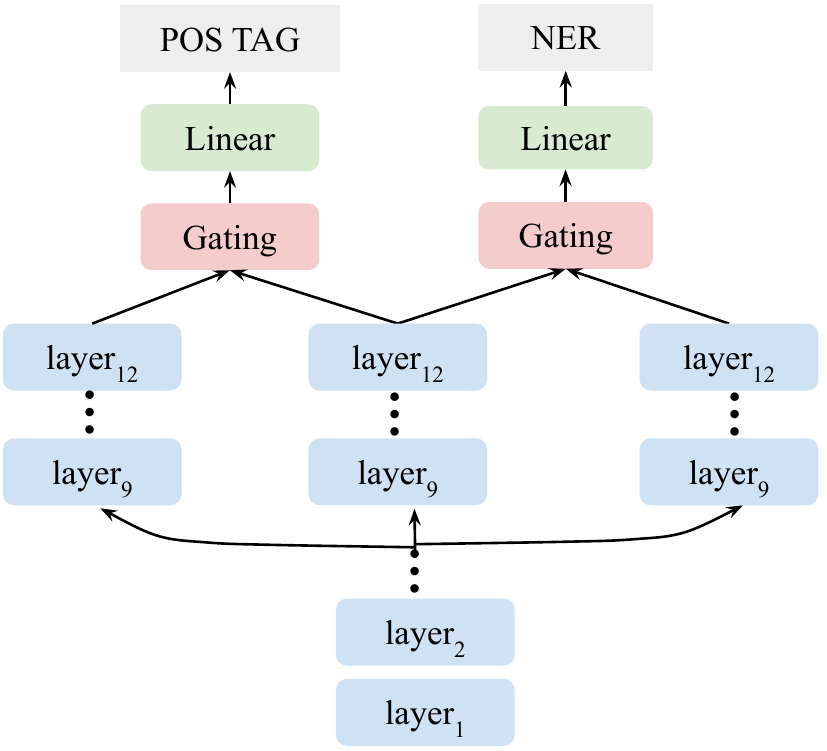}
    \caption{Model structure for multi-task active learning of the BERT base model. ``layer$_{\text{1},\cdots, \text{12}}$'' correspond to encoder layers in BERT-base encoder. We adopt the same structure following~\citet{rotman2022multi}.}
    \label{fig:model-structure}
\end{figure}

Following \citet{reichart2008multi,rotman2022multi}, we build upon BERT-base \citep{devlin-etal-2019-bert} for all experiments.
For each task, we add a linear layer on top of the BERT encoder.
For multi-task training, parameters in the BERT encoder are shared while parameters in the linear layers are not.
\Cref{fig:model-structure} shows an example of the model structure for the multi-task active learning setting.

\section{\sysname Results on CoNLL2003}
\label{appendix: conll2003-results}

\begin{figure*}
    \centering
    \subfigure[Named entity recognition]{\includegraphics[width=0.45\textwidth]{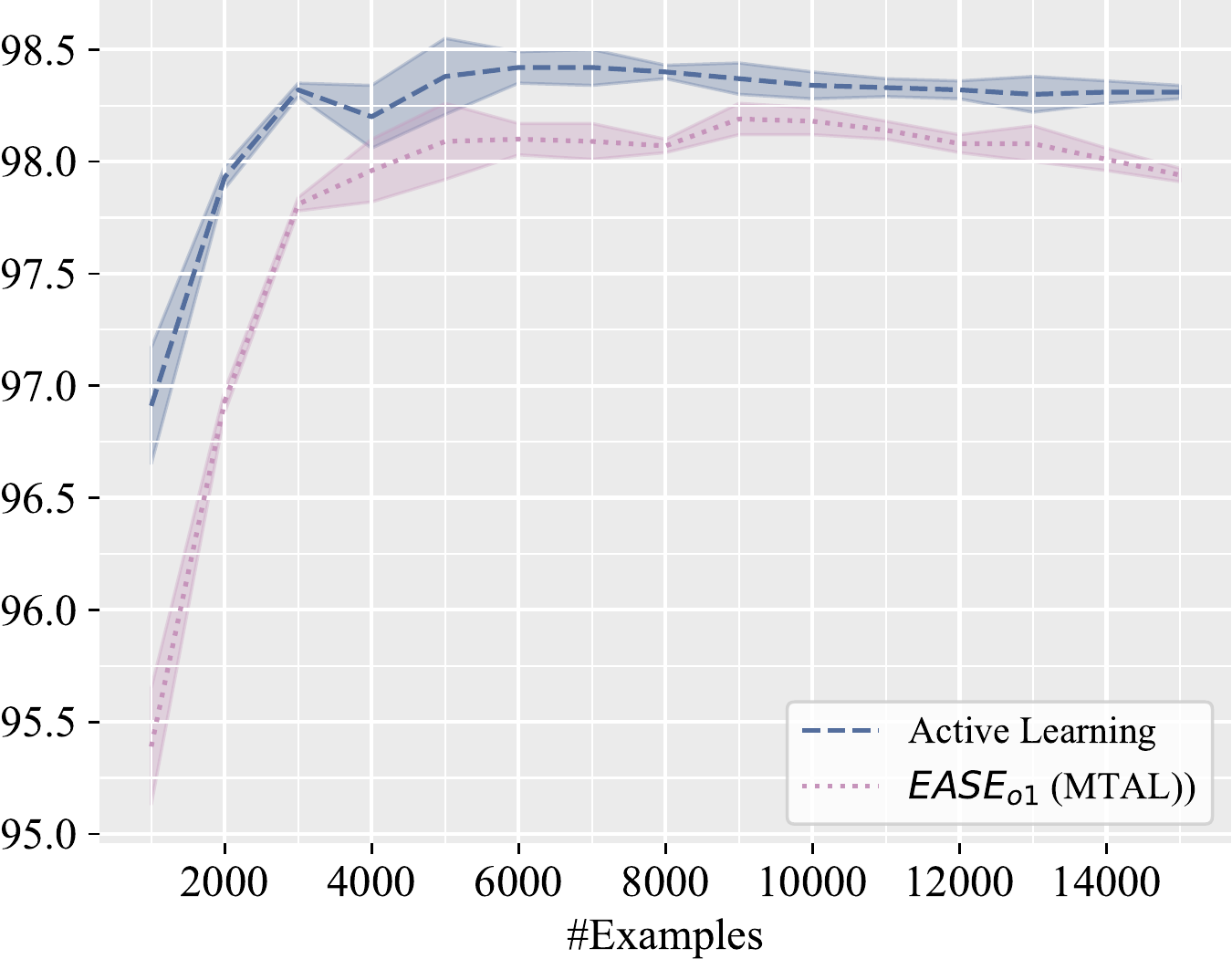}} 
    \subfigure[Chunk parsing]{\includegraphics[width=0.45\textwidth]{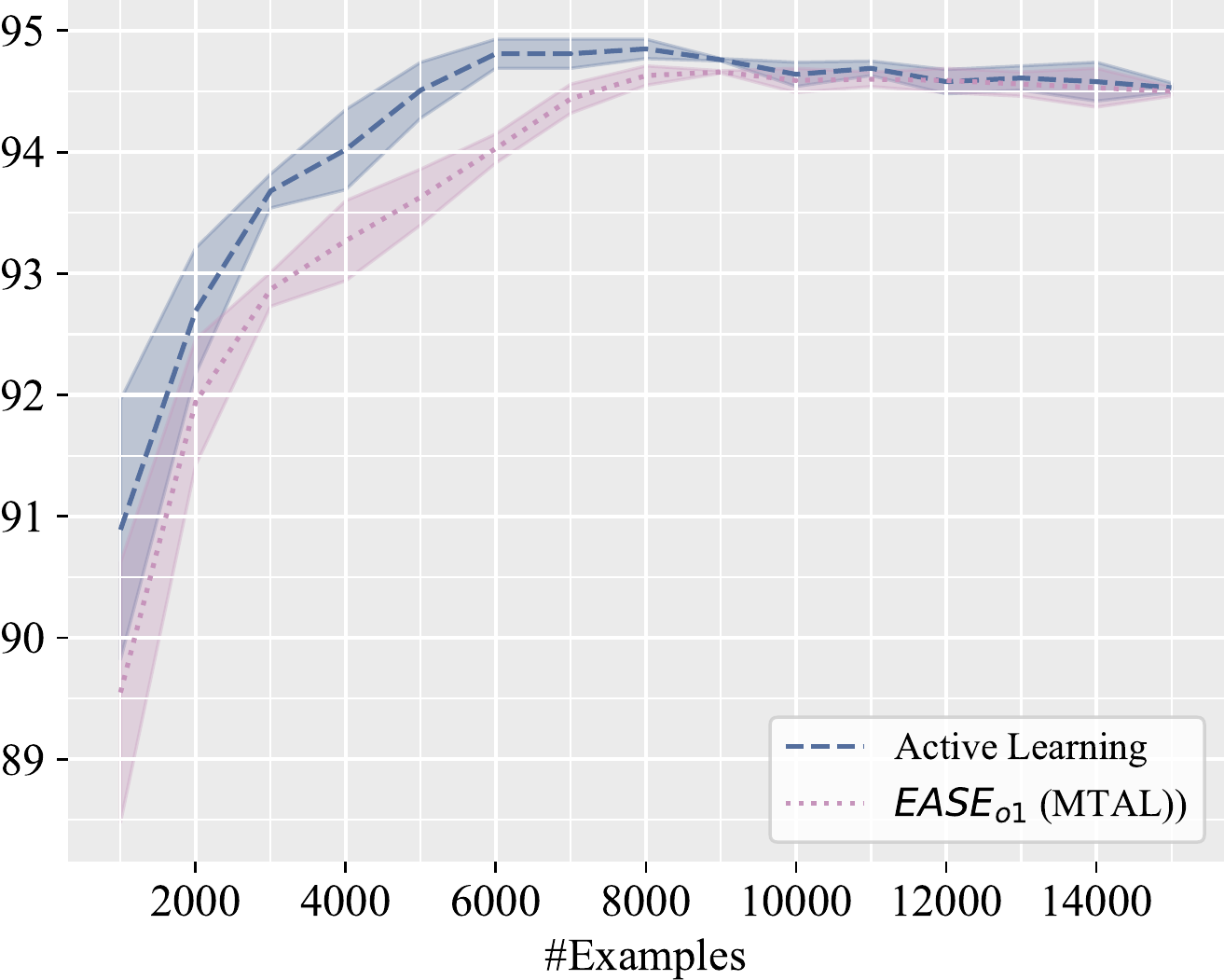}} 
    \caption{Validation accuracy v.s. the number of training instances for (a) named entity recognition and (b) chunk parsing for CoNLL2003 dataset.}
    \label{fig:conll2003-acc}
\end{figure*}

\begin{figure*}
    \centering
    \subfigure[Named entity recognition]{\includegraphics[width=0.45\textwidth]{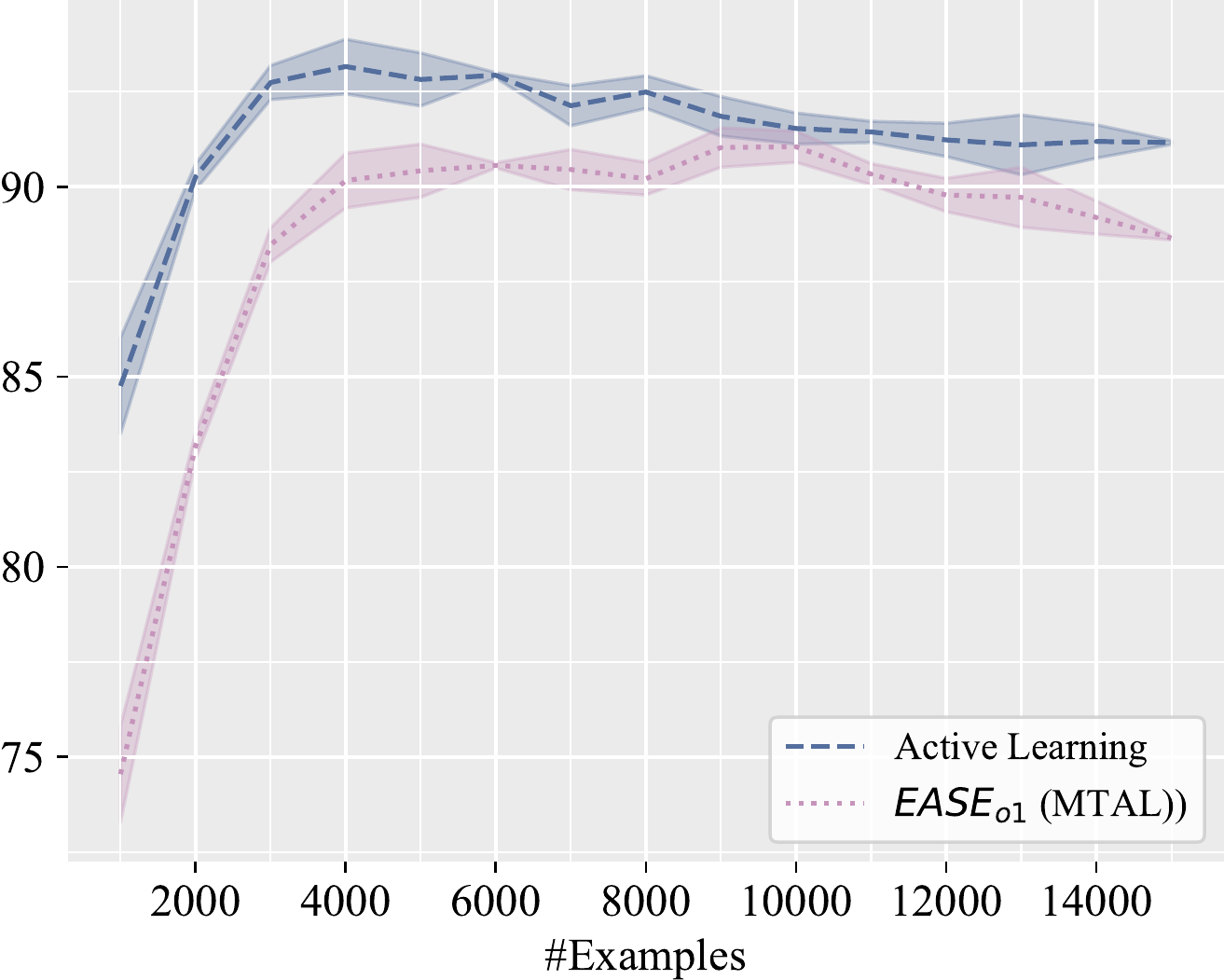}} 
    \subfigure[Chunk parsing]{\includegraphics[width=0.45\textwidth]{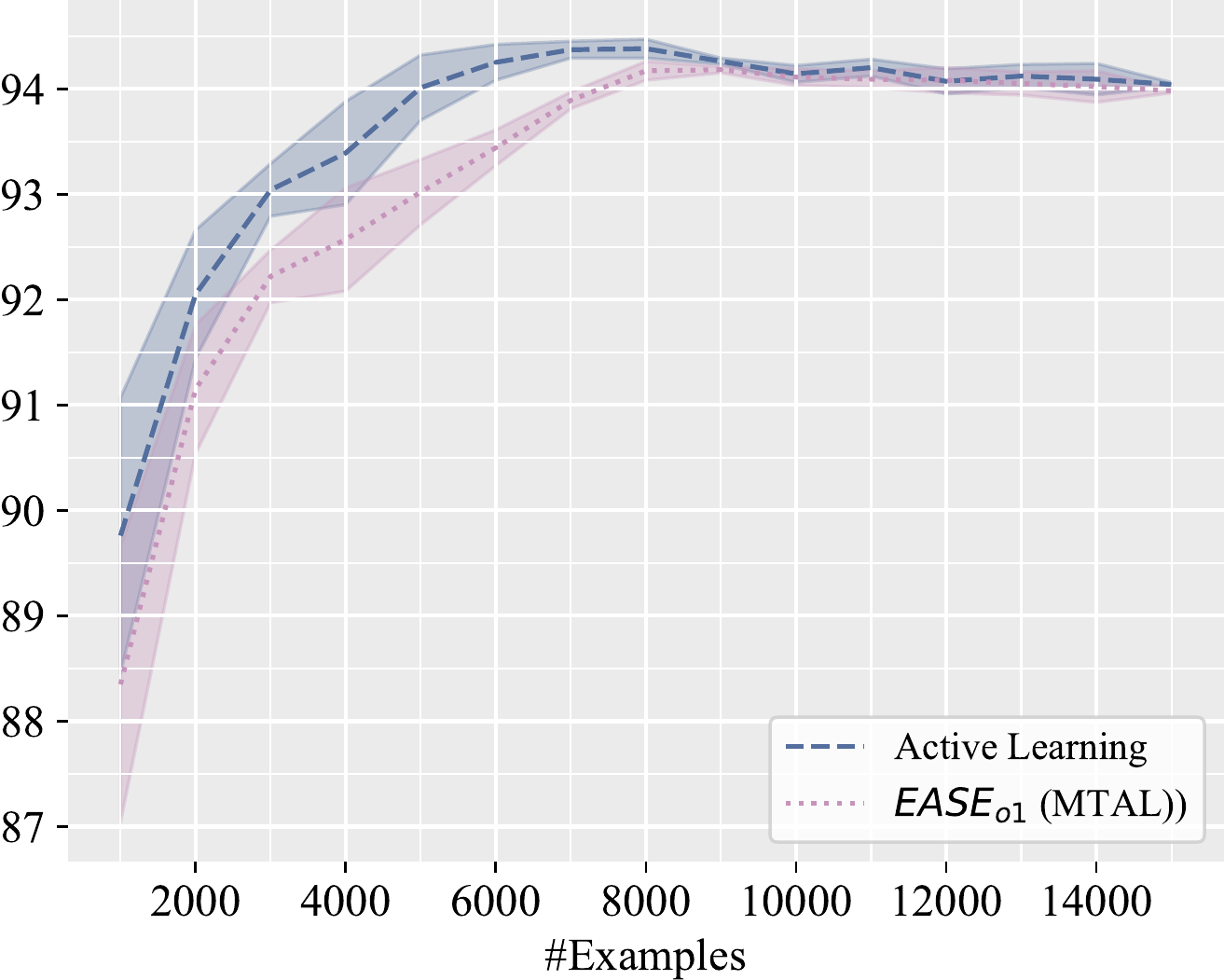}} 
    \caption{Validation recall v.s. the number of training instances for (a) named entity recognition and (b) chunk parsing for CoNLL2003 dataset.}
    \label{fig:conll2003-recall}
\end{figure*}

\begin{figure*}
    \centering
    \subfigure[Named entity recognition]{\includegraphics[width=0.45\textwidth]{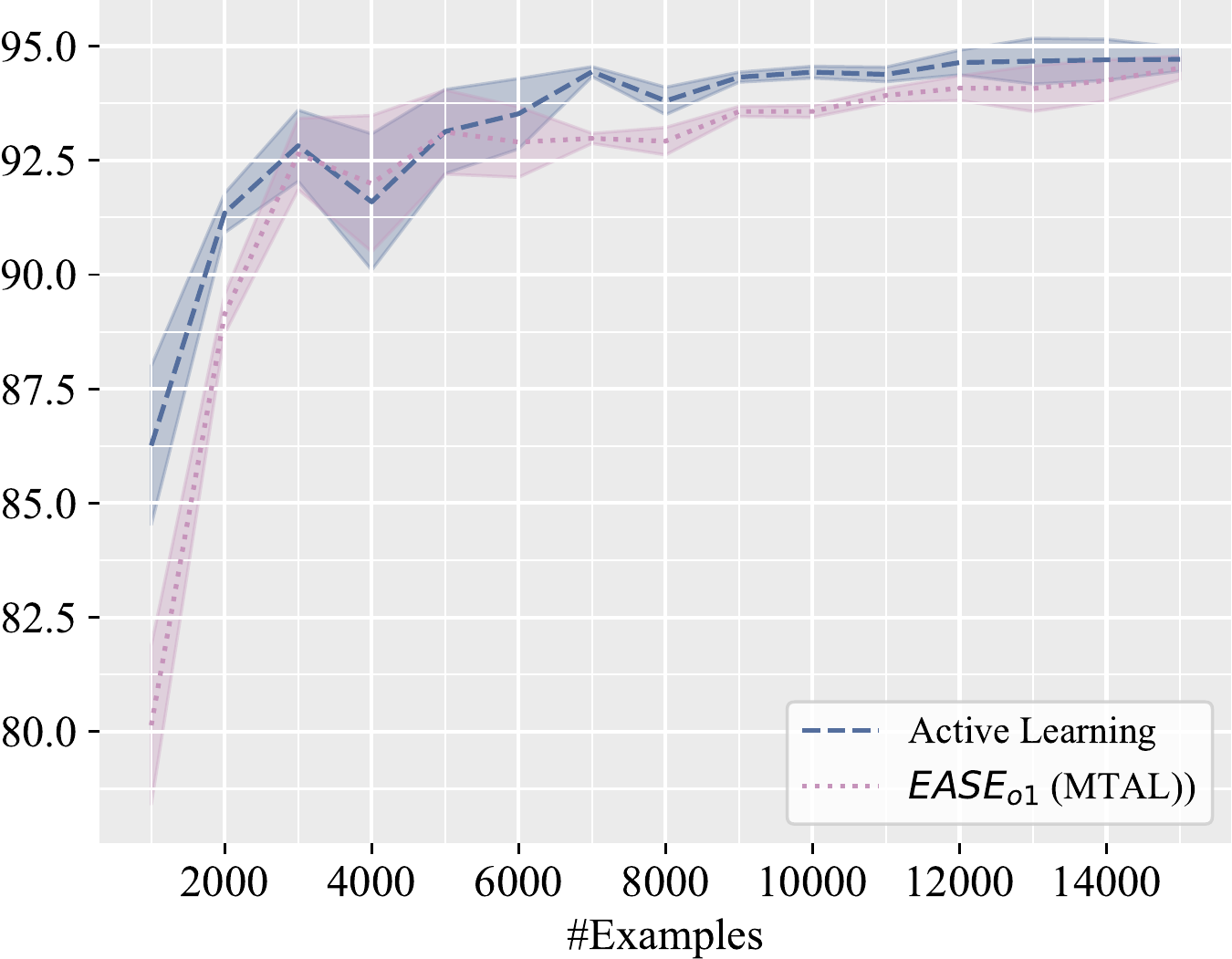}} 
    \subfigure[Chunk parsing]{\includegraphics[width=0.45\textwidth]{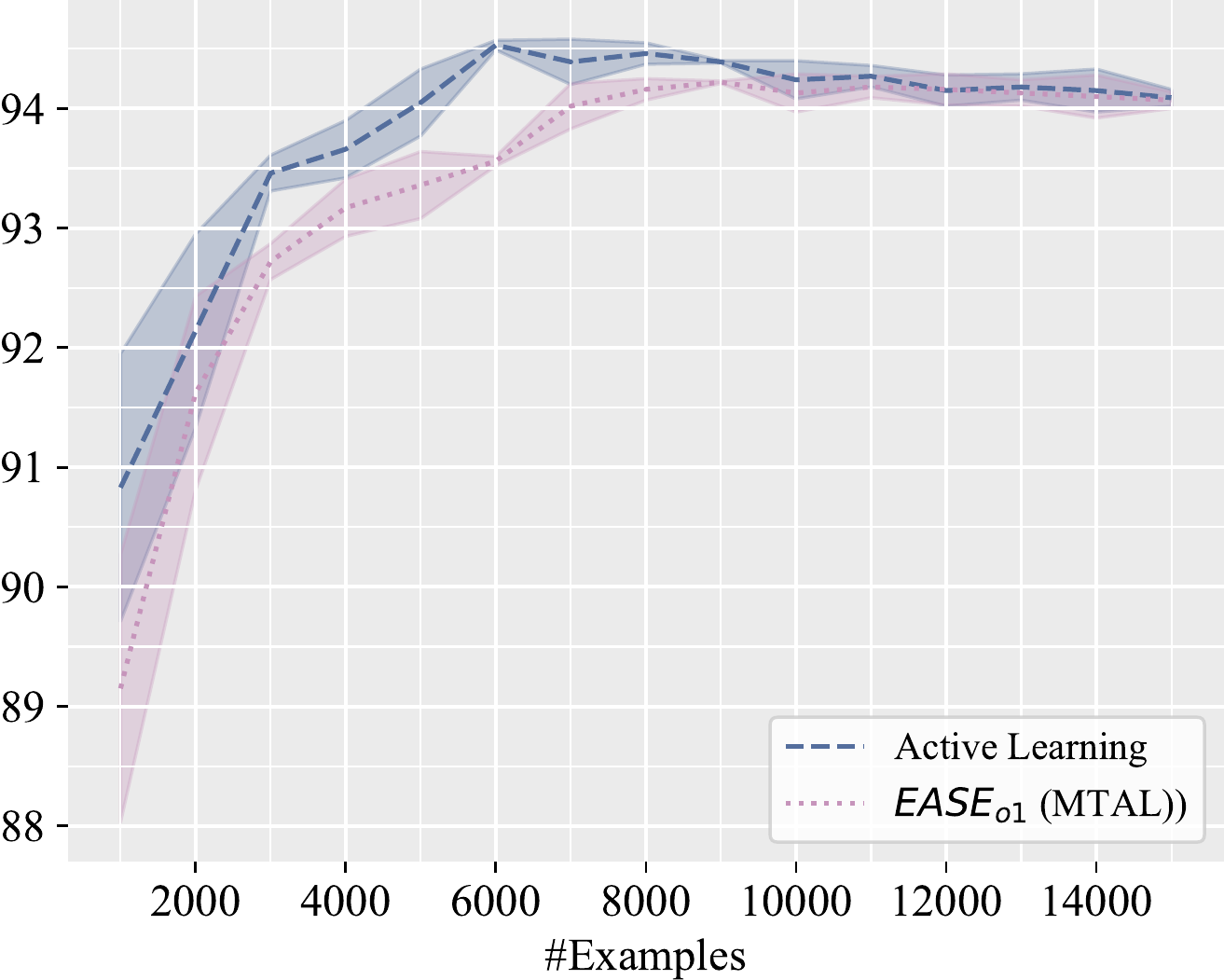}} 
    \caption{Validation precision v.s. the number of training instances for (a) named entity recognition and (b) chunk parsing on the CoNLL 2003 dataset.}
    \label{fig:conll2003-precision}
\end{figure*}

\begin{figure*}
    \centering
    \subfigure[Named entity recognition]{\includegraphics[width=0.45\textwidth]{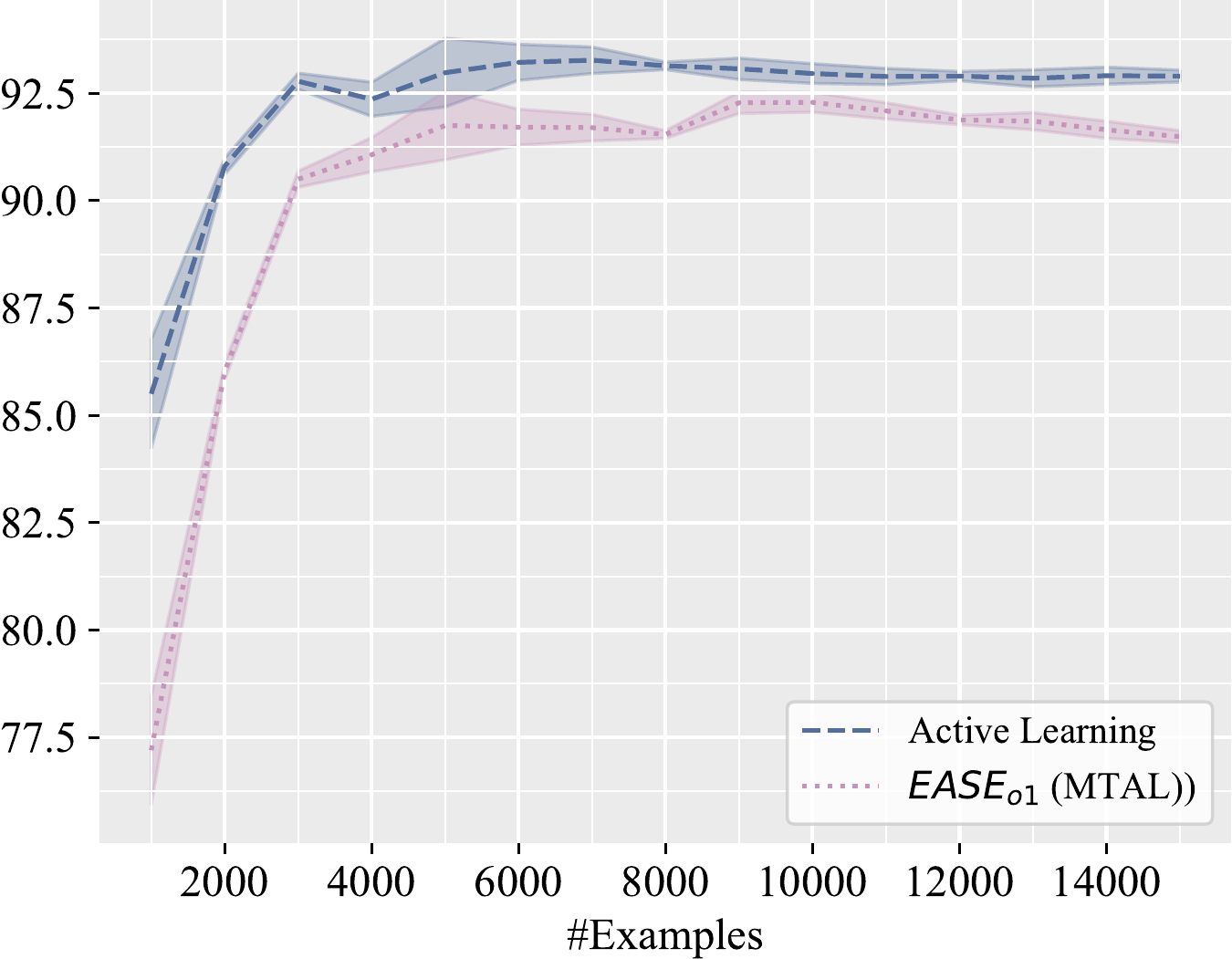}} 
    \subfigure[Chunk parsing]{\includegraphics[width=0.45\textwidth]{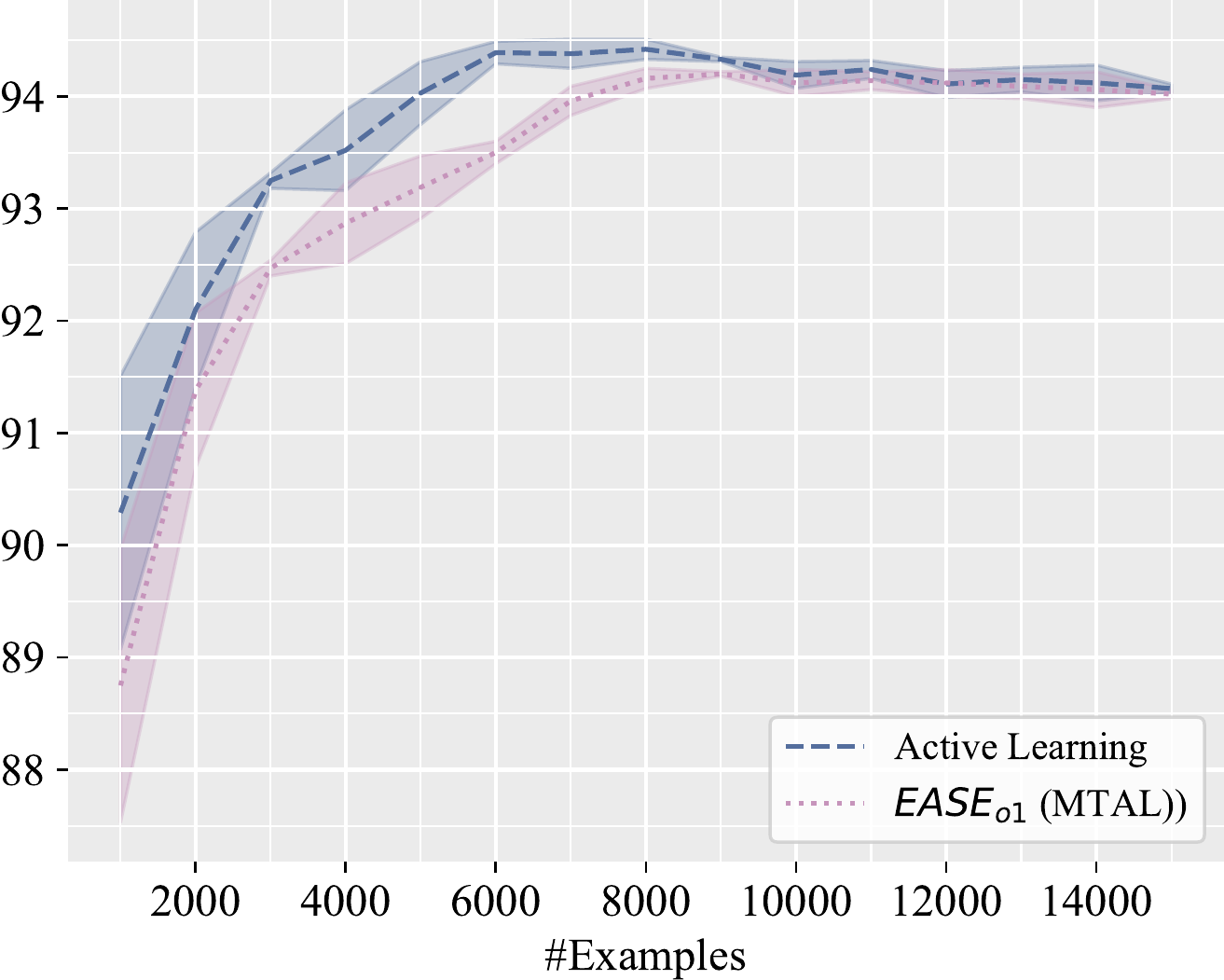}} 
    \caption{Validation f1 v.s. the number of training instances for (a) named entity recognition and (b) chunk parsing for CoNLL 2003 dataset.}
    \label{fig:conll2003-f1}
\end{figure*}

\Cref{fig:conll2003-acc,fig:conll2003-recall,fig:conll2003-precision,fig:conll2003-f1} show the model performance on the validation set as the training goes. In general \sysname$_{\text{o1}}$ (multi-task active learning) performs comparably with active learning.

\section{Experiment Set-ups for Back End Option 2: Active Learning With Demographic Features}
\label{appendix: demographic-experiment-set-ups}
The Sentiment Analysis dataset~\cite{diaz2018addressing} provides abundant demographic features such as age, race, gender, etc.
For each statement, annotators choose one of five sentiments: negative, somewhat negative, neutral, somewhat positive, or positive. 
We use the BERT-base model with a linear classification layer in the experiments. 
We experiment on various combinations of the demographic features in our preliminary experiments and find that the model trained with the age concatenated to the input performs the best.

\section{Prompt Examples}
\label{appendix: prompt-examples}

\Cref{fig:prompt} gives an example of the prompt for the task of named entity recognition.

\begin{figure}
    \centering
    \small
    \texttt{
Given the sentence `` 1860 Munich 3 1 0 2 3 5 3 '' the entities-recognition are `` B-ORG I-ORG O O O O O O O ''\textbackslash n\textbackslash n Given the sentence `` Production of cold laminates was 120,500 tonnes , 4.2 percent higher than the same month last year and 11 percent higher than in June . '' the entities-recognition are `` O O O O O O O O O O O O O O O O O O O O O O O O O ''\textbackslash n\textbackslash n Given the sentence `` Jul-18.Jul '' the entities-recognition are `` O ''\textbackslash n\textbackslash n Given the sentence `` 5/8 - Meluawati ( Indonesia ) beat Chan Chia Fong ( Malaysia ) 11-6 '' the entities-recognition are `` O O B-PER O B-LOC O O B-PER I-PER I-PER O B-LOC O O ''\textbackslash n\textbackslash n Given the sentence `` The poll , taken on Sunday and Monday as the president engaged in a whistle-stop train trip to the Democratic Convention in Chicago , put Clinton at 51 percent , Dole at 36 percent and Ross Perot of the Reform Party at 8 percent . '' the entities-recognition are `` O O O O O O O O O O O O O O O O O O O B-MISC I-MISC O B-LOC O O B-PER O O O O B-PER O O O O B-PER I-PER O O B-ORG I-ORG O O O O ''\textbackslash n\textbackslash n Given the sentence `` 1. Ilke Wyludda ( Germany ) 66.60 metres '' the entities-recognition are `` O B-PER I-PER O B-LOC O O O ''\textbackslash n\textbackslash n Given the sentence `` -DOCSTART- '' the entities-recognition are `` O ''\textbackslash n\textbackslash n Given the sentence `` Dynamo Kiev 5 Kremin Kremenchuk 0 '' the entities-recognition are `` B-ORG I-ORG O B-ORG I-ORG O ''\textbackslash n\textbackslash n Given the sentence `` MANCHESTER , England 1996-08-27 '' the entities-recognition are `` B-LOC O B-LOC O ''\textbackslash n\textbackslash n Given the sentence `` Sokol Tychy 5 Lech Poznan 3 '' the entities-recognition are `` B-ORG I-ORG O B-ORG I-ORG O ''\textbackslash n\textbackslash n Given the sentence `` Longyear is a town in mourning , a close-knit community that has been shattered . '' the entities-recognition are }
    \caption{Sample prompt given to the language model.}
    \label{fig:prompt}
\end{figure}

\section{Front-End User Study}
\label{appendix: front-end-user-study}

\subsection{Online Survey Content}

Here are the questions we asked in the online survey:

\begin{enumerate}
    \item Describe \textbf{one} of the NLP tasks that you want to work on (e.g. text-to-SQL, summarization, etc)
    \item Describe the data that you wish to gather for a specific task. 

Here are some examples for your reference:

\begin{enumerate}
    \item To investigate individual biases in summarization, data that can be collected includes individual summaries of a passage and the annotator's identification who created it.
    \item For analyzing models' performance on various question types in the Natural Language Inference task, the data collected could include text data in the form of questions along with their corresponding inferences in a text box. Additionally, annotators can be asked to categorize the question types by providing multiple-choice options for classification.
\end{enumerate}

\end{enumerate}

\subsection{Collected Answers}
\label{appendix subsec: front-end-user-answers}

We collected 10 answers from NLP researchers and list them in the form of \textbf{NLP task} and their detailed descriptions as follows:

\paragraph{Sequence editing} Given the source sentence, apply a set of actions to change the sentence. So the data will be (source sentence, set of actions, target sentence). Humans will need to apply several actions and the tool should be able to record those actions.

\paragraph{Counselor response editing} Provide an example of rewriting, then provide a response and ask for a rewrite. Importantly, I will also provide examples of varying degrees of content preservation, and ask the annotator to conform to the specified level.

\paragraph{Response generation} To investigate the cultural factors behind the conversations among people from different backgrounds, data that can be collected would include a conversation on a controversial topic and annotations on the speaker’s background information and the utterance aligned to those background annotations.

\paragraph{Commonsense inference} To study how annotators' cultural background affects the commonsense inference, the data can be collected by writing the tail entity given the head entity and relations or identifying inferences in the existing knowledge base that are incorrect in some cultures. 

\paragraph{Text to text generation} To investigate perspective differences of Tweets written by different languages, Tweets regarding a certain topic and their language labels can be collected.

\paragraph{Chinese sentiment analysis} I want the annotated Chinese texts. 

Generally, annotators should be asked to do multiple choices for different classifications like Positive, Neutral, and Negative. They may also be asked to label a range of sentiment levels in -3, -2, -1, 0, 1, 2, 3. Other necessary information includes AnnotatorID, Language background, Education background (CS, Linguistics, Chinese Studies, High Schools, etc.), and age.

In addition, the task of annotation could be a Chinese sentence. It could also be a Chinese paragraph. These two methods may result in two results.

If the research needs to compare Chinese sentiment annotations and English sentiment annotations, it would be better to shuffle the same meaning of texts in two languages. If the same meaning of texts in two languages appears at the same time, most annotators may choose the same evaluation.

\paragraph{Machine translation} I would want to have paired data for the source and target language, from a diverse range of sources to get good representations. Therefore, I want to collect translations for colloquial terms in daily dialogues.    

\paragraph{Commonsense reasoning QA in cultural contexts} Data could be collected in form of questions about events or actions together with a set of possible answers. Among those, one of the answers is the most plausible in the context described in the question.

\paragraph{Summarization} For evaluating existing models as well as formulating new ones to summarize dialogue, data can be collected by obtaining meeting transcripts paired with meeting summaries authored by participants of that meeting.

\paragraph{Dialogue evaluation} Given a dialogue, annotate the level of empathy across multiple dimensions. 

\subsection{Constructed Script}
\sysname can support all answers we collect in \Cref{appendix subsec: front-end-user-answers}. The JSON scripts to construct these interfaces are provided in our GitHub repo. 

\begin{figure}
    \centering
    \includegraphics[width=\linewidth]{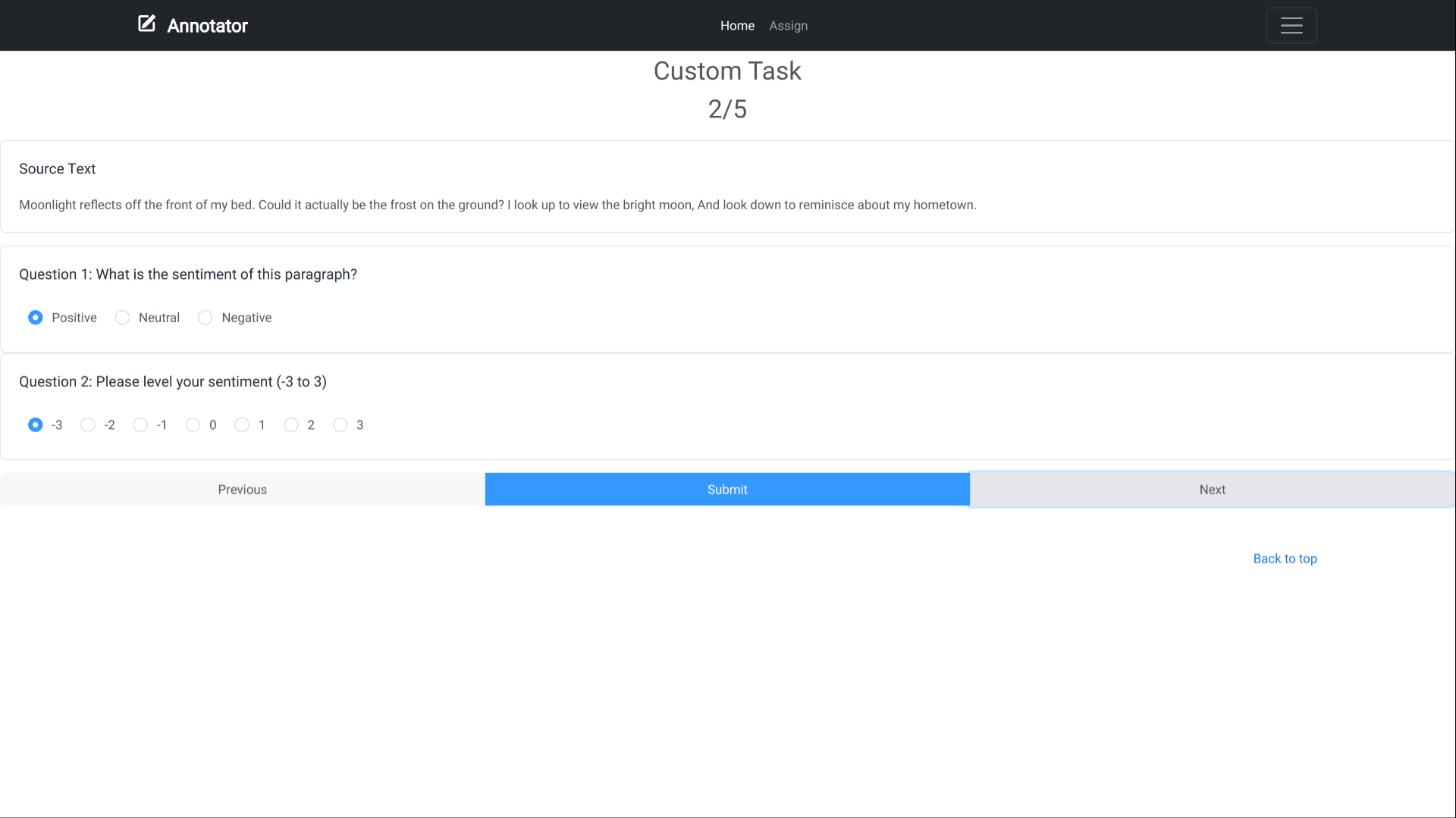}
    \caption{Example interface constructed for \textbf{Chinese sentiment analysis} by \sysname. For display purposes, we translate the Chinese poem into English.}
    \label{fig:user-study-poem}
\end{figure}

\Cref{fig:user-study-poem} provides the interface constructed for \textbf{Chinese sentiment analysis}.

Here we provide an example JSON script to render the annotation interface corresponding to \Cref{fig:user-study-poem} in Code Snippet~\ref{lst:json-chinese-sentiment-analysis} as the following:

\onecolumn
\begin{lstlisting}[language=json,firstnumber=1,title = {JSON file for the requested interface of Chinese sentiment analysis. For display purposes, we translate Chinese poems into English.}, captionpos=b, label={lst:json-chinese-sentiment-analysis}]
{
    "data": {
        "source": [
            "Goose, goose, goose,You bend your neck towards the sky and sing. Your white feathers float on the emerald water,Your red feet push the clear waves.",
            "Moonlight reflects off the front of my bed. Could it actually be the frost on the ground? I look up to view the bright moon, And look down to reminisce about my hometown.",
            "Green hills skirt the northern border, White waters gird the eastern town; Here we part with each other, And you set out like a lonesome wisp of grass, Floating across the miles, farther and farther away. You’ve longed to travel like roaming clouds, But our friendship, unwilling to wane as the sun is to set, Let it be here to stay. As we wave each other good-bye, Our horses neigh, as if for us they sigh."
        ],
        "question": [
            [
                "What is the sentiment of this paragraph?",
                "Please level your sentiment (-3 to 3)"
            ],
            [
                "What is the sentiment of this paragraph?",
                "Please level your sentiment (-3 to 3)"
            ],
            [
                "What is the sentiment of this paragraph?",
                "Please level your sentiment (-3 to 3)"
            ]
        ],
        "result": [
            [
                {
                    "result": 0
                },
                {
                    "result": 0
                }
            ],
            [
                {
                    "result": 0
                },
                {
                    "result": 0
                }
            ],
            [
                {
                    "result": 0
                },
                {
                    "result": 0
                }
            ]
        ],
        "done": [
            0,
            0,
            0
        ]
    },
    "format": [
        {
            "type": "button",
            "properties": {
                "contents": [
                    "Positive",
                    "Neutral",
                    "Negative"
                ]
            }
        },
        {
            "type": "button",
            "properties": {
                "contents": [
                    "-3",
                    "-2",
                    "-1",
                    "0",
                    "1",
                    "2",
                    "3"
                ]
            }
        }
    ]
}
% \end{lstlisting}

\twocolumn

\section{Example of JSON file}
\label{appendix: customized-json}
Code Snippet~\ref{lst:json-custom-interface} shows the JSON file that corresponds to the custom interface as follows:
\onecolumn
\begin{lstlisting}[language=json,firstnumber=1,title = {JSON file for the customized interface}, captionpos=b, label=lst:json-custom-interface]
{
    "data": {
        "source": [
            "The heat required for boiling the water and supplying the steam can be derived from various sources, most commonly from burning combustible materials with an appropriate supply of air in a closed space (called variously combustion chamber, firebox). In some cases the heat source is a nuclear reactor, geothermal energy, solar energy or waste heat from an internal combustion engine or industrial process. In the case of model or toy steam engines, the heat source can be an electric heating element.",
            "The most useful instrument for analyzing the performance of steam engines is the steam engine indicator. Early versions were in use by 1851, but the most successful indicator was developed for the high speed engine inventor and manufacturer Charles Porter by Charles Richard and exhibited at London Exhibition in 1862. The steam engine indicator traces on paper the pressure in the cylinder throughout the cycle, which can be used to spot various problems and calculate developed horsepower. It was routinely used by engineers, mechanics and insurance inspectors. The engine indicator can also be used on internal combustion engines. See image of indicator diagram below (in Types of motor units section)."
        ],
        "question": [
            [
                "What is the usual source of heat for boiling water in the steam engine?",
                "Aside from firebox, what is another name for the space in which combustible material is burned in the engine?",
                "What is the sentiment of this paragraph?",
                "Please do the part of speech tagging on this paragraph.",
                "Translate the paragraph into Chinese."
            ],
            [
                "What instrument is used to examine steam engine performance?",
                "What year saw the earliest recorded use of the steam engine indicator?",
                "What is the sentiment of this paragraph?",
                "Please do the part of speech tagging on this paragraph.",
                "Translate the paragraph into Chinese."
            ]
        ],
        "result": [
            [
                {
                    "result": []
                },
                {
                    "result": ""
                },
                {
                    "result": 0
                },
                {
                    "result": []
                },
                {
                    "result": ""
                }
            ],
            [
                {
                    "result": []
                },
                {
                    "result": ""
                },
                {
                    "result": 0
                },
                {
                    "result": []
                },
                {
                    "result": ""
                }
            ]
        ],
        "done": [
            0,
            0
        ]
    },
    "format": [
        {
            "type": "selection",
            "properties": {
                "contents": []
            }
        },
        {
            "type": "textbox",
            "properties": {}
        },
        {
            "type": "button",
            "properties": {
                "contents": [
                    "positive",
                    "negative",
                    "neutral"
                ]
            }
        },
        {
            "type": "selection",
            "properties": {
                "contents": [
                    "NP",
                    "PP",
                    "VP"
                ]
            }
        },
        {
            "type": "textbox",
            "properties": {}
        }
    ]
}
\end{lstlisting}

\twocolumn

\end{document}